\documentclass[acmsmall,screen,table]{acmart}
\setcopyright{acmcopyright}
\copyrightyear{2024}
\acmYear{2024}
\acmDOI{XXXXXXX.XXXXXXX}
\acmJournal{JACM}
\acmVolume{37}
\acmNumber{4}
\acmArticle{1}
\acmMonth{9}
\usepackage{pifont}

\usepackage{xcolor}
\usepackage{enumitem}
\usepackage{algorithmic}
\usepackage{graphicx}
\usepackage{textcomp}
\usepackage{booktabs}
\usepackage{xspace}
\usepackage{subcaption}
\usepackage{tcolorbox}
\usepackage{multirow}
\usepackage{listings}

\definecolor{highlightY}{HTML}{FAE6A2}

\definecolor{bg}{HTML}{F8F9FB}  
\definecolor{grey}{RGB}{224,224,224}

\usepackage[framemethod=TikZ]{mdframed}
\mdfdefinestyle{MyFrame}{%
    linecolor=black,
    outerlinewidth=0.3pt,
    roundcorner=5pt,
    skipabove = 5.5pt,
    skipbelow = 5.5pt,
    innertopmargin=5.5pt, 
    innerbottommargin=5.5pt, 
    innerrightmargin=5.5pt,
    innerleftmargin=5.5pt,
    backgroundcolor=bg, 
}

\begin{document}
\newcommand{\gp}{{\texttt{GooglePlay}}\xspace}
\newcommand{\gh}{{\texttt{GitHub}}\xspace}
\newcommand{\jr}{{\texttt{Jira}}\xspace}
\newcommand{\gr}{{\texttt{Gerrit}}\xspace}
\newcommand{\sof}{{\texttt{StackOverflow}}\xspace}
\newcommand{\llama}{{\textsc{Llama 2-Chat}}\xspace}
\newcommand{\wizardlm}{{\textsc{WizardLM}}\xspace}
\newcommand{\vicuna}{{\textsc{Vicuna}}\xspace}
\newcommand{\bert}{{\textsc{BERT}}\xspace}
\newcommand{\roberta}{{RoBERTa}\xspace}
\newcommand{\albert}{{\textsc{ALBERT}}\xspace}
\newcommand{\xlnet}{XLNet\xspace}
\newcommand{\db}{{{DistilBERT}}\xspace}
\newcommand{\xmarkg}{\textcolor{lightgray}{\ding{55}}\xspace}

\newboolean{showcomments}
\setboolean{showcomments}{true}
\ifthenelse{\boolean{showcomments}}
{ \newcommand{\mynote}[2]{
      \fbox{\bfseries\sffamily\scriptsize#1}
        {\small$\blacktriangleright$\textsf{\emph{#2}}$\blacktriangleleft$}}}
        { \newcommand{\mynote}[2]{}}
\newcommand{\todoc}[2]{{\textcolor{#1} {\textbf{#2}}}}
\newcommand{\todo}[1]{{\todoc{red}{\textbf{#1}}}}
\newcommand{\todored}[1]{\todoc{red}  {\textbf{#1}}}
\newcommand{\todoblue}[1]{\todoc{magenta}  {\textbf{#1}}}
\newcommand{\todopp}[1]{\todoc{black}  {\textbf{#1}}}

\newcommand{\ft}[1]{\mynote{Ferdian}{\todored{#1}}}
\newcommand{\zt}[1]{\mynote{Ting}{\todopp{#1}}}
\newcommand{\iv}[1]{\mynote{Ivana}{\todoblue{#1}}}

\title{Revisiting Sentiment Analysis for Software Engineering in the Era of Large Language Models}

\author{Ting Zhang}
\affiliation{%
  \institution{Singapore Management University}
  \country{Singapore}
}
\email{tingzhang.2019@phdcs.smu.edu.sg}
\orcid{0000-0002-6001-1372}

\author{Ivana Clairine Irsan}
\affiliation{%
    \institution{Singapore Management University}
  \country{Singapore}
}
\email{ivanairsan@smu.edu.sg}
\orcid{0000-0001-6350-2700}

\author{Ferdian Thung}
\affiliation{%
    \institution{Singapore Management University}
  \country{Singapore}
}
\email{ferdianthung@smu.edu.sg}
\orcid{0000-0002-5566-3819}

\author{David Lo}
\affiliation{%
  \institution{Singapore Management University}
  \country{Singapore}
}
\email{davidlo@smu.edu.sg}
\orcid{0000-0002-4367-7201}

\renewcommand{\shortauthors}{Zhang et al.}

\begin{abstract}
Software development involves collaborative interactions where stakeholders express opinions across various platforms. Recognizing the sentiments conveyed in these interactions is crucial for the effective development and ongoing maintenance of software systems. 
For software products, analyzing the sentiment of user feedback, e.g., reviews, comments, and forum posts can provide valuable insights into user satisfaction and areas for improvement. This can guide the development of future updates and features.
However, accurately identifying sentiments in software engineering datasets remains challenging.

This study investigates bigger large language models (bLLMs) in addressing the labeled data shortage that hampers fine-tuned smaller large language models (sLLMs) in software engineering tasks. We conduct a comprehensive empirical study using five established datasets to assess three open-source bLLMs in zero-shot and few-shot scenarios. Additionally, we compare them with fine-tuned sLLMs, using sLLMs to learn contextual embeddings of text from software platforms.

Our experimental findings demonstrate that bLLMs exhibit state-of-the-art performance on datasets marked by limited training data and imbalanced distributions. bLLMs can also achieve excellent performance under a zero-shot setting. However, when ample training data is available or the dataset exhibits a more balanced distribution, fine-tuned sLLMs can still achieve superior results.

\end{abstract}



\maketitle

\section{Introduction}
\label{sec:introduction}
Sentiment analysis (SA), or opinion mining, is the computational study of people's opinions or emotions toward entities~\cite{liu2012survey}.
SA holds significant practical value and has found diverse applications across a wide spectrum of domains, including but not limited to business, marketing, politics, healthcare, and public advocacy~\cite{drus2019sentiment}.
Generally speaking, SA contains several tasks, such as sentiment classification~\cite{liu2020sentiment}, aspect-based sentiment classification~\cite{zhang2022survey}, or hate speech detection~\cite{schmidt2017survey}.
Most existing Software Engineering (SE) research focuses on sentiment classification, i.e., assigning a sentiment polarity (e.g., negative, neutral, and positive) to a given text unit.
For simplicity, we refer to the sentiment classification as SA in the rest of this article.
SA has proven its utility in various SE tasks, exemplified by its role in evaluating user reviews of mobile applications~\cite{lin2018sentiment} and identifying sentences conveying negative opinions about application programming interfaces (APIs)~\cite{zhang2013extracting}.
Given that SE is inherently a collaborative endeavor, comprehending the sentiments expressed by different stakeholders across various platforms becomes imperative for the effective development and maintenance of software systems.

Prior studies have demonstrated that general SA tools work well on social media posts or product reviews while performing poorly on SE datasets~\cite{jongeling2015choosing,tourani2014monitoring}.
This discrepancy has spurred a growing interest in developing SE-specific SA tools over the past decade~\cite{ahmed2017senticr,chen2019sentimoji,islam2018sentistrength}.
These sentiment analysis for SE (SA4SE) tools usually either propose a SE-specific lexicon~\cite{islam2018sentistrength} or a SE-specific model~\cite{ahmed2017senticr}.
At the same time, several benchmarking studies on evaluating general SA tools and SE-specific tools have been conducted~\cite{lin2018sentiment,zhang2020sentiment,novielli2018benchmark,jongeling2015choosing}.
Zhang et al.~\cite{zhang2020sentiment} made the first attempt to embrace the power of language models, i.e., small pre-trained large language models (sLLMs),\footnote{To distinguish from the recent larger sizes of large language models, we consider \textbf{sLLMs} as relatively smaller sizes of large language models that can be easily fine-tuned locally, such as BERT, RoBERTa, and XLNet. sLLMs usually contain <1B parameters. We refer to the bigger large language models, which contain billions of parameters as \textbf{bLLMs}.} for SA4SE.
Zhang et al. demonstrated that sLLMs outperform existing specialized SA4SE tools on the evaluation datasets.
However, several challenges persist in the field of SA4SE.
First, the accuracy of sLLMs can degrade when there is a lack of labeled data for fine-tuning.
For instance, the \textcolor{black}{Google Play dataset~\cite{lin2018sentiment}}, which contains app review comments, only has 341 labeled documents and among them 25 are neutral ones.
The fine-tuned sLLMs predicted none of the data points in the test set as neutral.
While acquiring more labeled data can help mitigate this issue, manually labeling large volumes of data is time-consuming.
The second challenge relates to the limitations of fine-tuning itself. Fine-tuning sLLMs requires updating some of the model parameters with domain-specific data.
Lastly, the third challenge occurs in cross-platform settings~\cite{novielli2020can}, where SA4SE tools tend to perform poorly. Models trained from one dataset may not generalize well when tested on a different dataset, hindering the generalizability and effectiveness of existing SA4SE tools.
Given these challenges, there is a need to explore more effective solutions for SA4SE.

Recently, large language models have shown promising results in many areas, spanning from general natural language processing (NLP) tasks to specialized applications like software development.
bLLMs are usually trained on massive corpora of texts and contain many parameters.
For instance, GPT-3~\cite{brown2020language} contains 175 billion parameters.
\textsc{Llama}~\cite{touvron2023llama} is trained on trillions of tokens and contains 7B to 65B parameters.
Given the large number of parameters, fine-tuning bLLMs for every downstream task is impractical.
These bLLMs permit \textit{in-context learning}: they can be adapted to a downstream task simply by providing it with a prompt (a natural language description of the task)~\cite{bommasani2021opportunities}.
This adaptability has been a game-changer in reducing the need for domain-specific training data, as bLLMs can leverage their pre-existing knowledge to excel in diverse applications.
In-context learning has drastically reduced the domain-specific training examples required for a particular application~\cite{chowdhery2022palm}.
bLLMs can make predictions conditioned on a few input-output examples without updating model parameters and achieve success in various tasks~\cite{brown2020language,chowdhery2022palm}.
Nevertheless, their performance in SA4SE remains largely unexplored.
The intriguing prospect of adopting bLLMs in this context lies in their ability to potentially address the challenges associated with fine-tuning sLLMs and the limitations observed in cross-platform settings. 

To fill this gap, our work embarks on a journey to explore the effectiveness of bLLMs for SA4SE.
To investigate the effectiveness of bLLMs for SA4SE, we conducted a comprehensive empirical study on five existing SE datasets.
We first evaluate bLLMs under zero-shot and few-shot settings.
For the zero-shot setting, we experimented with three different prompt templates.
For the few-shot setting, we experimented with 1-, 3-, and 5-shot.
The experimental results demonstrate that bLLMs can perform well under a zero-shot setting, while few-shot learning can further boost the performance.
However, adding more shots does not guarantee an improvement in the performance.
We also compared prompting bLLMs with fine-tuning sLLMs.
We find that, on the dataset that lacks training data and the data is highly imbalanced, bLLMs can surpass sLLMs by a large gap.
For the datasets that contain sufficient training data and more balanced data, sLLMs may still be preferred.

Our contribution can be summarized as follows:
\begin{itemize}
    \item{Our work is the first study to examine the effectiveness of open-source bLLMs on the SA4SE task.}
    \item{We evaluate three open-source bLLMs under zero and few-shot settings.}
    \item{We compare fine-tuned \textcolor{black}{sLLMs} with bLLMs on five SE datasets collected from five distinct platforms.}
\end{itemize}

The remaining parts of this work are as follows:
Section~\ref{sec:preliminaries} introduces the background of our work.
Section~\ref{sec:setup} discusses about our experimental setup.
Section~\ref{sec:results} presents the results of our empirical study.
Section~\ref{sec:discussion} discusses the additional experiments, the implications of our findings, and the threats to validity.
Section~\ref{sec:conclusion} concludes this work and discusses future work.
\section{Related Work}
\label{sec:preliminaries}
In recent years, SA4SE has emerged as a vibrant and active research area within SE. 
LLMs also have been widely applied in the SE domain. This section provides an overview of the relevant literature, primarily focusing on the techniques proposed to boost SA4SE accuracy and empirical studies in SA4SE.

\subsection{Boosting SA4SE Accuracy}
In the past decades, many techniques have been proposed to improve the effectiveness of identifying sentiments or emotions in the SE domain~\cite{chen2019sentimoji,chen2021emoji,islam2018sentistrength,imran2022data,ahmed2017senticr,calefato2018sentiment,calefato2019emtk,islam2019marvalous,murgia2018exploratory,islam2018deva,biswas2020achieving}.

Chen et al.~\cite{chen2019sentimoji} propose SEntiMoji, an emoji-powered learning approach for SA in SE. They employ emotional emojis as noisy labels of sentiments and propose a representation-learning approach that uses both Tweets and GitHub posts containing emojis to learn sentiment-aware representations for SE-related texts.
In the evaluation, they compare SEntiMoji with four SA4SE tools on sentiment polarity benchmark datasets.
The experimental results show that SEntiMoji can significantly improve the performance.

Furthermore, Chen et al.~\cite{chen2021emoji} include an additional evaluation of SEntiMoji on the emotion detection task.
They also compared it with four existing emotion detection methods, including DEVA~\cite{islam2018deva}, EmoTxt~\cite{calefato2019emtk}, MarValous~\cite{islam2019marvalous}, and ESEM-E~\cite{murgia2018exploratory}.
The experimental results on the five benchmark datasets covering 10,096 samples for sentiment detection and four benchmark datasets covering 10,595 samples for emotion detection demonstrate that SEntiMoji is effective.

Besides developing a new SA4SE tool, \textcolor{black}{another research line aims} at improving SA accuracy by handling existing challenges, such as labeled data scarcity. 
Imran et al.~\cite{imran2022data} address the data scarcity problem by automatically creating new training data using a data augmentation technique. 
They specifically target the data augmentation strategy to improve the performance of emotion recognition by analyzing the types of errors made by popular SE-specific emotion recognition tools.
Their results show that when trained with their best augmentation strategy, three existing emotion classification tools, i.e., ESEM-E, EMTk, and SEntiMoji, received an average improvement of 9.3\% in micro-F1 score.

As previously discussed in Section~\ref{sec:introduction}, Zhang et al.~\cite{zhang2020sentiment} introduced the approach of fine-tuning sLLMs for SA4SE.
Our approach distinguishes itself from the aforementioned research by pioneering the use of prompting bLLMs for SA4SE. Alongside these efforts, we are collectively working towards advancing the accurate identification of sentiment within the SE field.

\subsection{Empirical Studies in SA4SE}
With the proliferation of domain-specific SA4SE tools, a series of empirical investigations have been conducted to illuminate our understanding of this field's progress and challenges~\cite{lin2018sentiment,novielli2018benchmark,jongeling2015choosing,obaidi2022sentiment,jongeling2017negative,novielli2021assessment,lin2022opinion}.

Novielli et al., in their recent study~\cite{novielli2021assessment}, delve into the critical question of how off-the-shelf, SE-specific SA tools affect the conclusion validity of empirical studies in SE. They begin by replicating two prior studies that explore the role of sentiment in security discussions on GitHub and question-writing on Stack Overflow. Subsequently, they extend these studies by assessing the level of agreement between different SA tools and manual annotations, using a gold standard dataset comprising 600 documents. The experimental findings from this research reveal that when applied out-of-the-box, various SA4SE tools may yield conflicting results at a granular level. Consequently, it becomes imperative to consider platform-specific fine-tuning or retraining to account for differences in platform conventions, jargon, or document lengths.

Obaidi et al.~\cite{obaidi2022sentiment} conducted a systematic mapping study to comprehensively examine SA tools developed for or applied in the SE domain. This study summarizes insights drawn from 106 papers published up to December 2020, focusing on six key aspects: (1) the application domain, (2) the purpose of SA, (3) the datasets used, (4) the approaches for developing SA tools, (5) the utilization of pre-existing tools, and (6) the challenges faced by researchers. Based on their findings, neural networks emerge as the top-performing approach, with BERT identified as the most effective tool.

Beyond the scope of sentiment classification, some researchers have explored broader facets of SA4SE, such as opinion mining. Opinion mining encompasses a wider spectrum of tasks than the sentiment polarity identification typically evaluated in SA4SE studies. It includes SA, subjectivity detection, opinion identification, and joint topic SA. In a comprehensive systematic literature review, Lin et al. investigated 185 papers on opinion mining in Software Engineering~\cite{lin2022opinion}, shedding light on the diverse research efforts in this area.

Each of these existing empirical studies has contributed valuable insights into the evolving landscape of SA4SE. Notably, our work stands apart from these studies as it introduces the use of bLLMs to this domain for the first time.
\section{Experimental Setup}
\label{sec:setup}

\subsection{Research Questions}
\begin{figure}[t]
\centering
\includegraphics[width=\textwidth]{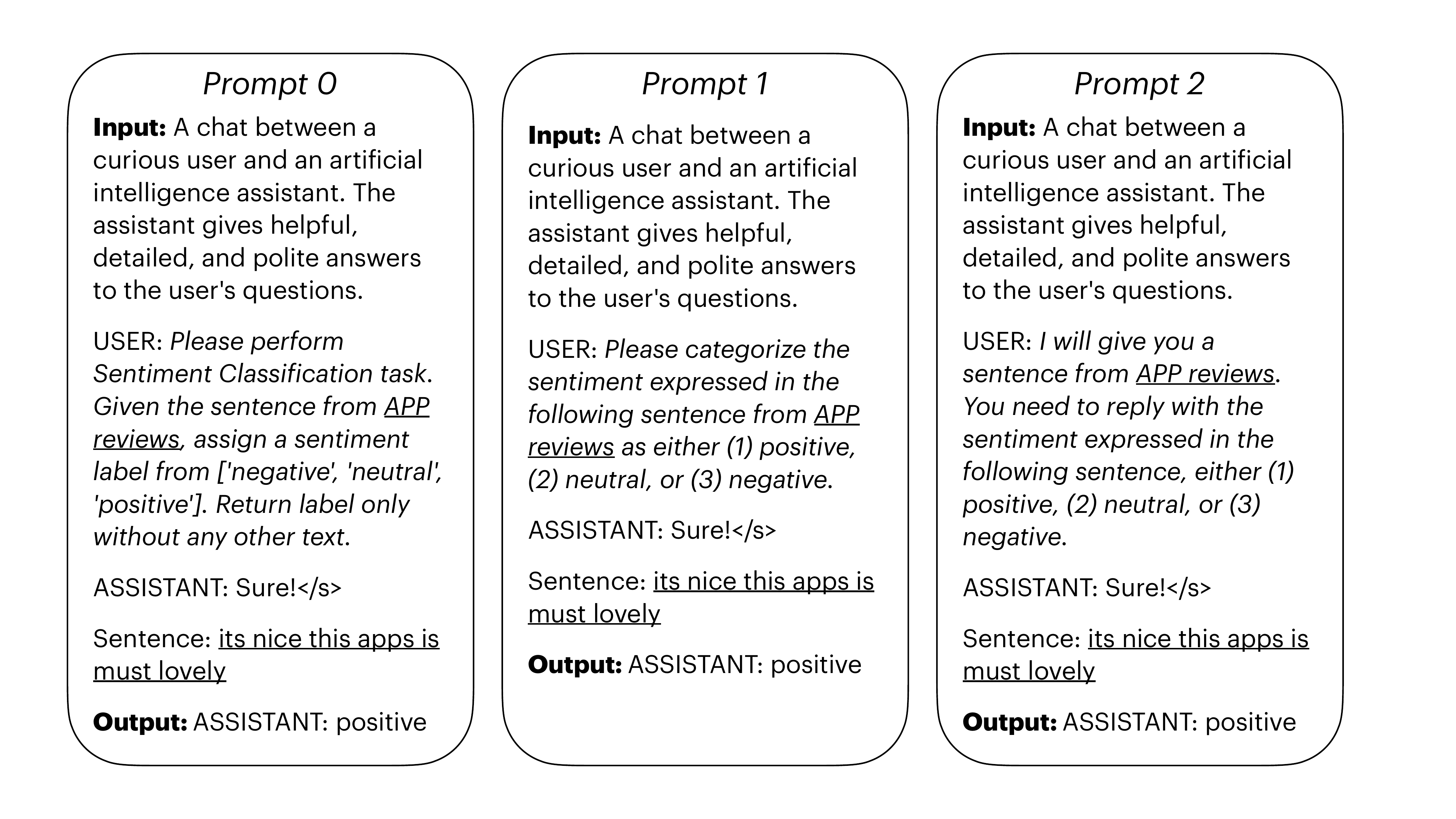}
\caption{The zero-shot prompt templates we utilized when running \vicuna~\cite{vicuna2023} and \wizardlm~\cite{xu2023wizardlm}}
\label{fig:zero-shot-prompt}
\end{figure}

In this work, we plan to answer the following research questions (RQs):

\begin{itemize}
\item[RQ1:] \textbf{How do various prompts affect the performance of bLLMs in zero-shot learning for the SA4SE task?} \newline 
In this RQ, our initial focus is exploring the zero-shot learning scenario, where bLLMs are prompted without providing any labeled data. 
Prior studies have unveiled that bLLMs exhibit varying results even when prompted with semantically similar queries~\cite{perez2021true,lu2022fantastically,mizrahi2023state}. 
Additionally, \textcolor{black}{earlier} research findings have emphasized the substantial impact of different word orders within the prompt templates on the predictions~\cite{min2022rethinking,deng2023llms}.
\textcolor{black}{Nevertheless, it is yet to be determined if bLLMs exhibit varying performance when prompted with templates that have subtle differences on the SA4SE task.}
\textcolor{black}{We experiment with slight variations of the prompts to precisely measure the bLLMs' sensitivity to nuanced linguistic cues. This approach is essential in controlled experimental settings where isolating the impact of single variables is crucial. By altering the prompts minimally, we can more accurately attribute any observed differences in performance to these specific changes, thereby gaining insight into how the model processes and interprets instructions at a granular level. Additionally, we believe the commands used to conduct SA are rather straightforward, making it challenging to create two prompts with substantial changes.}
\textcolor{black}{To evaluate the effectiveness of different prompt templates in generating accurate predictions, we hypothesize that there are significant differences in performance metrics (macro-F1, micro-F1, and AUC scores) among the various prompt templates. Specifically, we propose the following hypotheses for pairwise comparisons:}
\begin{itemize}
    \item \textcolor{black}{Prompt 0 vs. Prompt 1}
    \begin{itemize}
        \item \textcolor{black}{H0.1.1 (Null Hypothesis): There is no significant difference in the performance metrics between Prompt 0 and Prompt 1.}
        \item \textcolor{black}{H1.1.1 (Alternative Hypothesis): There is a significant difference in the performance metrics between Prompt 0 and Prompt 1.}
    \end{itemize}
    \item \textcolor{black}{Prompt 0 vs. Prompt 2}
    \begin{itemize}
        \item \textcolor{black}{H0.1.2 (Null Hypothesis): There is no significant difference in the performance metrics between Prompt 0 and Prompt 2.}
        \item \textcolor{black}{H1.1.2 (Alternative Hypothesis): There is a significant difference in the performance metrics between Prompt 0 and Prompt 2.}
    \end{itemize}
    \item \textcolor{black}{Prompt 1 vs. Prompt 2}
    \begin{itemize}
        \item \textcolor{black}{H0.1.3 (Null Hypothesis): There is no significant difference in the performance metrics between Prompt 1 and Prompt 2.}
        \item \textcolor{black}{H1.1.3 (Alternative Hypothesis): There is a significant difference in the performance metrics between Prompt 1 and Prompt 2.}
    \end{itemize}
\end{itemize}
We will employ a Wilcoxon signed-rank test with Bonferroni correction for multiple comparisons to assess the significance of the differences between these prompt templates. The corrected alpha threshold for significance will be set at 0.0167 to account for the multiple pairwise comparisons.

\item[RQ2:]  \textbf{How do various shots affect the performance of bLLMs in few-shot learning for the SA4SE task?} \newline
\textcolor{black}{In this RQ, we evaluate whether bLLMs exhibit enhanced performance in scenarios with a limited number of examples available. In contrast to zero-shot learning, few-shot learning incorporates an extra ``Demonstration'' component. While previous studies, such as Zhang et al.~\cite{zhang2023sentiment}, indicate that few-shot learning may surpass zero-shot learning in certain aspects, contrasting findings by Reynolds et al.~\cite{reynolds2021prompt} suggest that zero-shot prompts can significantly outperform few-shot prompts in some NLP tasks.
Nonetheless, it is still unclear whether bLLMs derive greater benefits from few-shot learning, specifically in the SA4SE task.}
\textcolor{black}{We aim to test the following null hypothesis: H0.2: there is no significant difference in the results obtained under few-shot learning compared to those from zero-shot learning.}

\item[RQ3:] \textbf{How do prompting bLLMs compare with fine-tuned sLLMs on the SA4SE task?} \newline
\textcolor{black}{Prior work has shown that fine-tuning sLLMs can achieve state-of-the-art results on the SA4SE task.
However, the effectiveness of prompting bLLMs compared to fine-tuning sLLMs remains an open question. }
To address this, we compare the best macro-F1 and micro-F1 scores achieved through prompting bLLMs with those from fine-tuned sLLMs.
\textcolor{black}{To further explore this comparison, we test the null hypothesis: H0.3: There is no significant difference in results obtained from prompting bLLMs and fine-tuning sLLMs.}

\item[RQ4:] \textcolor{black}{\textbf{What are the factors leading to the misclassification of sentiment labels by bLLMs?}} \newline
\textcolor{black}{To gain a deeper understanding of the reasons behind incorrect sentiment label classifications by bLLMs in the context of software engineering, we intend to carry out a detailed manual analysis.}
\end{itemize}

\begin{figure}[t]
  \centering
  \includegraphics[width=\textwidth]{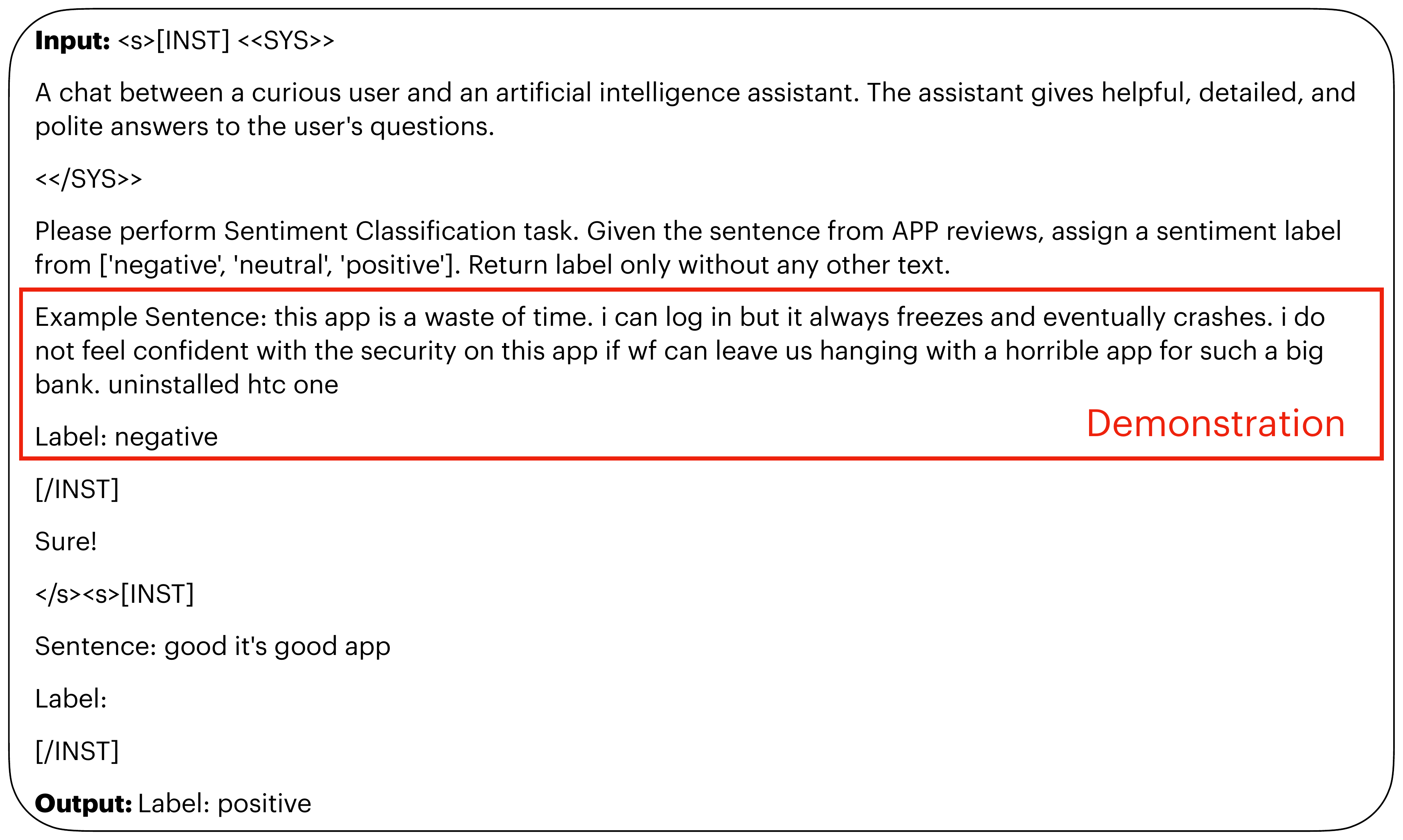}
  \caption{Few-shot prompt template (with $k=1$) utilized by \llama~\cite{touvron2023llama2}.}
  \label{fig:few-shot-prompt}
\end{figure}

\subsection{Method}
\textcolor{black}{
In this section, we present the steps performed to answer our research questions.
}

\textcolor{black}{
For \textbf{RQ1}, we investigate how various prompt templates affect the results. 
Given the straightforward nature of SA, our objective is to formulate equally straightforward prompts.
These prompts encompass two key components: the \textit{Task Description} and the \textit{Output Format}. The \textit{Task Description} serves the purpose of elucidating the task clearly and concisely. 
In our specific context, the task pertains to SA, and we articulate it through various expressions within the three templates. 
Importantly, the sentence origin (e.g., from APP reviews, from Stack Overflow)
remains consistent across all prompt templates, enabling the distinction of diverse contexts and domains.
The \textit{Output Format} component is designed to provide bLLMs with guidance for generating responses in a specific format, facilitating sentiment label extraction.
To maintain generality, we employ an identical prompt template for all five datasets.\footnote{There is only a minor difference in the \jr dataset, given it only contains two sentiments, i.e., negative and positive. We reduced the scope to only two options in the templates used by \jr.} 
Figure~\ref{fig:zero-shot-prompt} shows the three prompt templates we used for zero-shot setting.
While all prompt templates share a semantic similarity, they differ in their syntactic structure. 
}

\textcolor{black}{
Our inspiration for the first prompt template (i.e., Prompt 0) draws from Zhang et al.~\cite{zhang2023sentiment}. 
They designed the prompt to include only essential components, namely the task name, task definition, and output format.
In addition to these components, considering the specific context of SE which differs from a general context, we have incorporated the sentence origin. 
This addition aims to enrich the context provided to the bLLMs, enhancing their effectiveness in this specialized field.
For Prompt 1 and Prompt 2, we intentionally implemented subtle variations, focusing specifically on the aspects of \textit{structure} and \textit{tone}.
Regarding the \textit{structure} of the prompts, Prompt 0 stands out as the most formal and task-oriented.
In contrast, Prompts 1 and 2 introduce sentiment categories in an option-based format (1, 2, 3), rather than a straightforward list.
This choice gives these prompts a more interactive and engaging feel, making them especially suitable for situations that require user interaction. In addition, Prompt 2 presents a more interactive scenario where the user is actively involved (``I will give you a sentence...''). This format implies a back-and-forth interaction.
In terms of \textit{tone}, Prompt 1 is slightly more engaging than Prompt 0, as it reads more like a request than a command.
Prompt 2 further amplifies this interactive quality by establishing a dialogue-like contex.
}

\textcolor{black}{For \textbf{RQ2}, in the context of few-shot learning, we leverage the best-performing zero-shot prompt template, namely \textit{Prompt 0}.
We enrich \textit{Prompt 0} with various numbers of examples filled in the \textbf{Demonstration} part.
The demonstration part encompasses $k$ ($k=1,3,5$) examples and corresponding ground-truth labels, adhering to the desired format.
Figure~\ref{fig:few-shot-prompt} illustrates the few-shot prompt template employed by \llama on the \gp dataset. 
In the depicted figure, the demonstration segment (enclosed within the red box) comprises only one example. In the case of a 3-shot or 5-shot setup, this demonstration section would encompass a greater number of example sentences and their corresponding gold labels. We systematically sampled 1-, 3-, and 5-examples from the training data of each dataset, subsequently populating the demonstration segment within the template. This approach ensures that under the $k$-shot setting, different bLLMs receive the same set of examples.}

\textcolor{black}{For \textbf{RQ3}, we aim to compare prompting bLLMs and fine-tuning sLLMs. Thus, we ensure that both bLLMs and sLLMs make predictions on the same test data set.
We report the highest macro- and micro-F1 scores achieved by the bLLM on the dataset as the result achieved by the bLLM.
}

\textcolor{black}{In \textbf{RQ4}, we conduct a quantitative and qualitative analysis to understand the main cause of misclassification made by bLLMs.
For the quantitative aspect, we focus on the results from the most effective templates. 
Specifically, we analyze the results from \textit{Prompt 0} for zero-shot learning and from the \textit{5-shot} prompt for few-shot learning.
In the qualitative part of our study, we utilize the error categorization framework established by Novielli et al.~\cite{novielli2018benchmark,novielli2020can}.
\textcolor{black}{Employing these predefined categories, our analysis involved two authors (from here on, evaluators) independently assigning a category to each document in our study.
It is important to note that not all categories identified by Novielli et al. were necessarily covered in each document. 
Subsequently, the evaluators engaged in discussions to resolve any conflict and reach a consensus.
}
}

\subsection{Dataset}
\label{sec:dataset}
In this work, we experimented with the existing manually labeled datasets from five distinct platforms: Gerrit, GitHub, Google Play, Jira, and Stack Overflow. For simplicity, we refer to these datasets using abbreviations: \textcolor{black}{\gr~\cite{ahmed2017senticr}, \gh~\cite{novielli2020can}, \gp~\cite{lin2018sentiment}, \jr~\cite{lin2018sentiment}, and \sof~\cite{lin2018sentiment}}.

\vspace{6px}

\noindent{\textbf{\gr Dataset}: Ahmed et al.~\cite{ahmed2017senticr} meticulously labeled this dataset. They initiated their process by mining code review repositories from 20 prominent open-source software (OSS) projects. Three raters individually labeled the selected code review comments and resolved conflicts through discussion. The dataset was refined into two classes: \textit{negative} and \textit{non-negative}, forming the final dataset.}

\vspace{6px}

\noindent{\textbf{\gh Dataset}: Novielli et al.~\cite{novielli2020can} curated the \gh dataset, which comprises pull request and commit comments. Sentiment was assessed based on the entire comment, rather than isolated portions. The labeling process began with the manual classification of 4,000 comments, followed by a semi-automatic approach using Senti4SD~\cite{calefato2018sentiment}, which required manual confirmation of the automatically assigned polarity labels.}

\vspace{6px}

\noindent{\textbf{\gp, \jr, and \sof Datasets}: Lin et al.~\cite{lin2018sentiment} provided three datasets, each with its unique characteristics:

\begin{itemize}
    \item{\textbf{\gp Dataset}: Originally collected by Chen et al.~\cite{chen2015simapp}, this dataset contains user reviews of Android apps on Google Play. Villarroel et al.~\cite{villarroel2016release} selected a subset of reviews from Chen et al.'s dataset, and Lin et al. further sampled 341 reviews. Lin et al. performed the manual labeling of sentiment, where two annotators individually classified text as positive, neutral, or negative. In cases of disagreement, a third evaluator was involved for resolution.}

    \item{\textbf{\jr Dataset}: This dataset comprises Jira issue comment sentences and was originally collected and labeled by Ortu et al.~\cite{ortu2016emotional}. However, Ortu et al.'s dataset only provided emotional labels, such as \textit{love}, \textit{joy}, \textit{anger}, and \textit{sadness}. Lin et al. mapped sentences labeled with \textit{love} or \textit{joy} to ``positive'' and those labeled with \textit{anger} or \textit{sadness} to ``negative''.}

    \item{\textbf{\sof Dataset}: Lin et al. gathered and labeled the \sof dataset, extracting 5,073,452 sentences from the latest Stack Overflow dump available in July 2017. The sentences were selected based on two criteria: they had to be tagged with    ``Java'', and they needed to contain keywords such as ``library/libraries'' or ``API(s)''. A random sample of 1,500 sentences was manually labeled by assigning a sentiment score to each sentence. Two annotators carried out the manual annotation individually, with conflicts resolved through discussion.}
\end{itemize}
}

We split each dataset with a ratio of 8:1:1, which stands for training, validation, and test, respectively.
We did a stratified split where we kept the original class distribution in training, validation, and test.
Since running bLLMs is expensive, they usually contain billions of parameters, we did a sampling on all the test data with a confidence level of 95\% and a margin of error of +/- 5\%.\footnote{We included all the provided test data in the \gp dataset, as the number of sampled data is only 2 data points fewer than the whole test data.}
Similarly, we also kept the class distribution the same as in the original whole dataset.
\begin{table}[t]
\centering
\caption{Dataset statistics. Neg. stands for negative, Neu. stands for neutral, Pos. stands for positive, and Non-neg. stands for non-negative.}
\label{tab:dataset}
{
\begin{tabular}{llrrr}
\toprule
\multirow{2}{*}{\bf Dataset} & \multirow{2}{*}{\bf Total} & \textbf{Test} & \textbf{Sampled} & {\bf Avg.}\\
 & & & \textbf{Test} & {\bf Tokens} \\
\midrule
{\bf \gr} & 1,600: 398 (Neg.) 1,202 (Non-neg.) & 160 & 114 & 29 \\
{\bf \gh} & 7,122: 2,087 (Neg.)  3,022 (Neu.) 2,013 (Pos.) & 713 & 250 & 19 \\
{\bf \gp} & 341: 130 (Neg.) 25 (Neu.) 186 (Pos.) & 35 & 35 & 27\\
{\bf \jr} & 926: 636 (Neg.) 290 (Pos.) & 93 & 76  & 9\\
{\bf \sof} & 1,500: 178 (Neg.) 1,191 (Neu.) 131 (Pos.) & 150 & 109 & 11 \\
\bottomrule
\end{tabular}
}
\end{table}

Table~\ref{tab:dataset} presents the statistics of investigated datasets, specifically the average number of tokens per document. Notably, \gr and \gp exhibit longer text, likely due to the inclusion of code review and APP review comments, which often span multiple sentences. Although the \gh dataset also comprises pull request and commit comments, they are typically short in nature. 
Conversely, the \sof dataset and \jr dataset include \textit{sentences} from Stack Overflow and Jira, respectively. 
While some document units in these datasets contain multiple sentences and others just one, we collectively refer to them as \textit{documents}. \textcolor{black}{Note that the \gr, \gp, \jr, and \sof datasets investigated in our study are imbalanced, while the \gh dataset is more balanced.}

\subsection{\textcolor{black}{Evaluated} Language Models}
\label{sec:evaluated_lms}
\noindent{\textbf{bLLMs.} We include three recently proposed bLLMs based on their performance in the MMLU benchmark on the chatbot leaderboard~\footnote{\url{https://huggingface.co/spaces/lmsys/chatbot-arena-leaderboard}} in August 2023; the model name in the parenthesis is the exact model variant we used on the Hugging Face platform~\cite{wolf2020transformers}.} 

\begin{itemize}
    \item{\textbf{\llama} (\texttt{meta-llama/Llama-2-13b-chat-hf})~\cite{touvron2023llama2} is a fine-tuned version of \textsc{Llama 2} that is optimized for dialogue use cases. \textsc{Llama 2} uses the standard Transformer architecture~\cite{vaswani2017attention} and it applies pre-normalization with RMSNorm~\cite{zhang2022opt}, the SwiGLU activation function~\cite{shazeer2020glu}, and rotary positional embeddings~\cite{su2021roformer}. \textsc{Llama 2} made several improvements over \textsc{Llama 1}, including but not limited to more robust data cleaning, trained on 40\% more total tokens, and doubled the context length.}
    \item{\textbf{\vicuna} (\texttt{lmsys/vicuna-13b-v1.5})~\cite{vicuna2023} is a chatbot trained by fine-tuning \textsc{Llama 2} on 70K user-shared ChatGPT conversations. To better handle multi-turn conversations and long sequences, \vicuna is trained with the enhanced training script from Alpaca~\cite{alpaca}.}
    \item{\textbf{\wizardlm} (\texttt{WizardLM/WizardLM-13B-V1.2})~\cite{xu2023wizardlm} is another fine-tuned version of \textsc{Llama 2}. The authors propose \textit{Evol-Instruct}, which is a novel method using bLLMs instead of humans to automatically mass-produce open-domain instructions of various complexity levels to improve the performance of bLLMs. The resulting bLLMs by fine-tuning \textsc{Llama 2} with the evolved instructions is called \wizardlm.}
\end{itemize}

\vspace{8px}
\noindent{\textbf{sLLMs.} We include all the four sLLMs \textcolor{black}{evaluated by} Zhang et al.~\cite{zhang2020sentiment}, i.e., \bert~\cite{devlin2018bert}, \roberta~\cite{liu2019roberta}, \albert~\cite{lan2020albert}, \xlnet~\cite{yang2019xlnet}.
In addition, we also include a lightweight and memory-efficient variant of BERT, i.e., \db. 
We briefly describe these models.
They mainly differ in the pre-training tasks adopted.
We also present the exact model we used on the Hugging Face platform~\cite{wolf2020transformers} in the parenthesis.
\begin{itemize}
    \item{\textbf{\bert} (\texttt{bert-base-uncased})~\cite{devlin2018bert}, which stands for Bidirectional Encoder Representations from Transformers, introduces two key pre-training tasks. The first is mask language modeling (MLM), where BERT learns to predict masked words in a given text. Additionally, BERT incorporates the next sentence prediction (NSP) task, training to determine whether the second sentence logically follows the first or is a random sentence from the training data.}
    \item{\textbf{\roberta} (\texttt{roberta-base})~\cite{liu2019roberta} is short for ``A Robustly Optimized BERT Pretraining Approach''. RoBERTa is a BERT variant distinguished by its innovative training strategies and hyperparameter choices. Notably, it eliminates the NSP task, employs a larger batch size, trains on a larger corpus than BERT, and utilizes a dynamic masking strategy during training.}
    \item{\textbf{\albert} (\texttt{albert-base-v2})~\cite{lan2020albert}, or ``A Lite BERT'', is another BERT variant designed to reduce model size and computational requirements while maintaining or improving performance. ALBERT retains the MLM task but replaces the NSP task with the sentence order prediction (SOP) task. In SOP, ALBERT is trained to predict whether pairs of sentences are correctly ordered or if their positions have been swapped.}
    \item{\textbf{\xlnet} (\texttt{xlnet-base-cased})~\cite{yang2019xlnet} primarily focuses on capturing contextual information and long-range dependencies in text. It employs an autoregressive pretraining method and introduces permutation language modeling, where word order in a sentence is randomly shuffled, and the model is trained to predict the original sequence. XLNet also incorporates innovations such as the ``two-stream self-attention'' mechanism.}
    \item{\textbf{\db} (\texttt{distilbert-base-uncased})~\cite{sanh2019distilbert} is a distilled and smaller version of the BERT model. \db is designed to be faster and more memory-efficient. \db adopts model compression or knowledge distillation to learn from a teacher BERT to capture the same knowledge but with fewer parameters. As \db combines efficiency and strong performance, it has been popular in research and industry settings.}
\end{itemize}
}

\subsection{Evaluation Metrics}
Following prior works~\cite{novielli2018benchmark,zhang2020sentiment,novielli2020can}, we also report macro- and micro-averaged precision, recall, and F1-score.

\begin{align}
\label{formula-macro-p}
    Precision = \frac{\text{TP}}{\text{TP} + \text{FP}}
\end{align}

\begin{align}
\label{formula-macro-r}
    Recall = \frac{\text{TP}}{\text{TP} + \text{FN}}
\end{align}

\begin{align}
\label{formula-macro-f1}
    F1 = \frac{2 \times (Precision \times Recall)}{Precision + Recall}
\end{align}

\begin{align}
\label{formula-macro-f1-all}
    Macro\text{-}F1=\frac{F1_{negative}+F1_{neutral}+F1_{positive}}{3}
\end{align}

\vspace{8px}
\noindent{\textbf{Macro-averaged metrics:}} To get the macro-averaged metrics, we calculate each class and find their unweighted mean.
In our context, if we have three labels, i.e., negative, natural, and positive, we separately calculate the F1 for each class (See Formula~\ref{formula-macro-p}-\ref{formula-macro-f1}).
Macro-F1 would be the average of the F1 for these three classes (See Formula~\ref{formula-macro-f1-all}).

\vspace{8px}
\noindent{\textbf{Micro-averaged metrics:}} Formulate~\ref{formula-micro-p},~\ref{formula-micro-r}, and ~\ref{formula-micro-f1} shows how to calculate $micro\text{-}precision$, $micro\text{-}recall$, and $micro\text{-}F1$.
\begin{align}
\label{formula-micro-p}
    Micro\text{-}Precision = \frac{\text{Total TP}}{\text{Total TP} + \text{Total FP}}
\end{align}
\begin{align}
\label{formula-micro-r}
    Micro\text{-}Recall = \frac{\text{Total TP}}{\text{Total TP} + \text{Total FN}}
\end{align}
\begin{align}
\label{formula-micro-f1}
    Micro\text{-}F1 = \frac{2 \times (Micro\text{-}Precision \times Micro\text{-}Recall)}{Micro\text{-}Precision + Micro\text{-}Recall}
\end{align}

\vspace{8px}
Recall that $TP$, $FP$, and $FN$ are short for the true positives (i.e., where the model correctly predicts a positive class, and it matches the ground truth), false positives (i.e., where the model incorrectly predicts a positive class, but the ground truth is actually a negative class), and false negatives (i.e., where the model incorrectly predicts a negative class, but the ground truth is actually a positive class).
``Total TP'' represents the sum of true positives, ``Total FP'' represents the sum of false positives, and ``Total FN'' represents the sum of false negatives over all classes.

We report both the macro-F1 and micro-F1 scores as they show the balance between precision and recall.
We attach the full result in our replication package.\footnote{\url{https://github.com/soarsmu/LLM4SA4SE}}
From these formulas, we can find that micro-F1 emphasizes overall accuracy, while macro-F1 gives equal weight to each class's performance.
As we do not give weights to the classes and want to consider the F1 in each class, we choose macro-F1 as the main metric.
This choice of preferring macro-F1 also aligns with the prior work~\cite{novielli2020can}.

\vspace{8px}
\noindent{\textcolor{black}{\textbf{AUC:} Other than the widely used F1 score, we also report the area under the ROC curve (AUC) to evaluate the performance of the models.  AUC measures the entire two-dimensional area underneath the entire ROC curve. AUC is a popular metric for binary classification tasks, measuring the model's ability to distinguish between positive and negative classes. AUC values range from 0 to 1, with higher values indicating better performance. Some of our datasets have three classes, so we use the one-vs-rest strategy to compute the AUC of each class against the rest~\cite{fawcett2006introduction}.
}}

\subsection{Implementation Details}
Note that we only fine-tune sLLMs and prompt bLLMs, while (1) not fine-tuning bLLMs as bLLMs contain a large number of parameters, it is expensive to fine-tune them; and (2) not prompting sLLMs as they usually are pre-trained with the MLM task mentioned before. They predict missing words in a sentence, which differs from the autoregressive nature of models like GPT-3 that generate text sequentially based on a prompt. This architectural difference makes it less straightforward to use BERT for prompt-style tasks.
\textcolor{black}{sLLMs, like BERT, were designed for understanding existing text, not generating new text sequences. For example, if we input the prompt ``this app is a waste of time...'' (as shown in Figure~\ref{fig:few-shot-prompt}) into BERT, it will produce contextualized representations for each token but will not generate a sentiment token or new text. BERT's pre-training objective focused on representing input text, not text generation. To adapt BERT for sentiment classification, we fine-tune it on a labeled dataset to predict sentiment labels (positive, negative, neutral) based on the input text. During fine-tuning, the model learns to leverage BERT's contextualized representations for accurate predictions but does not learn to generate new sequences. Hence, we fine-tune sLLMs for classification tasks like sentiment analysis, not text generation. In contrast, bLLMs like \llama are pre-trained for text generation by predicting the next token based on context, making them better suited for generating coherent text sequences from prompts.
}

\lstset{
    basicstyle=\footnotesize,
    backgroundcolor=\color{bg},
    frame=single,
    framesep=5pt,
    xleftmargin=0.5ex,
    breaklines=true,
    escapeinside=||,
    keywordstyle=\color{black}, 
    stringstyle=\color{green}, 
    commentstyle=\color{gray}, 
}

\vspace{8px}
\noindent{\textbf{Prompting bLLMs.}} We use heuristics to extract the sentiment returned by bLLMs.
When bLLMs consider it hard to decide the sentiment (they replied with a sentiment other than the three polarities, e.g., ``mixed''), we label the predicted sentiment as ``neutral''.
As there is no ``neutral'' sentiment in the \jr dataset, if any bLLM predicts the sentiment as ``neutral'', we label the predicted sentiment as the opposite of the ground-truth label.
In the few-shot setting, we select $k$ examples (``shots'') at random from the training set ($(k=\{1,3,5\})$), and for each example, we append its ground-truth label.
To make a new prediction for a new example, we append one sentence from the test set.

\begin{figure}[t]
    \centering
    \includegraphics[width=\textwidth]{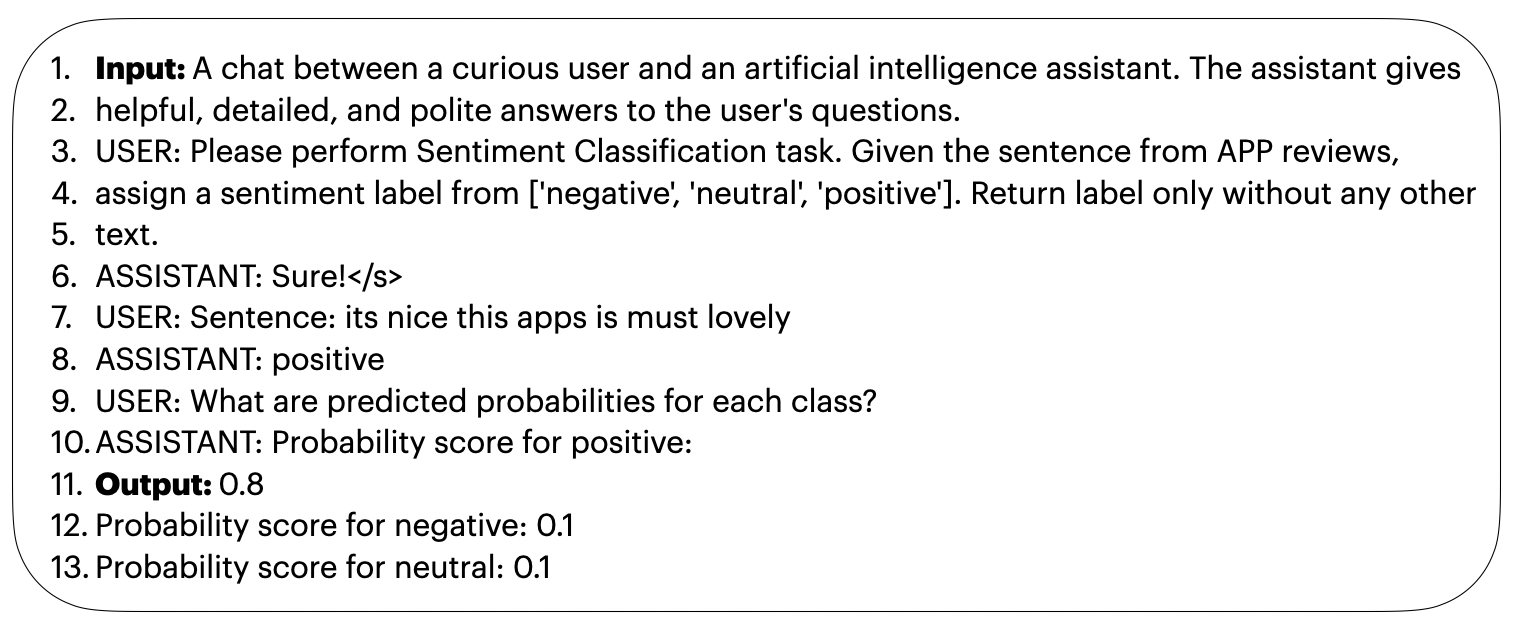}
    \caption{One example to get the prediction probability scores from the bLLMs.}
    \label{fig:example_auc}
\end{figure}

\textcolor{black}{
It is important to note that bLLMs do not directly provide probability scores for each class. To obtain these scores, we use the predicted result as a new prompt and ask the model to give the probability scores for each class. For example, Figure~\ref{fig:example_auc} illustrates an instance using the \vicuna model with Prompt 0 in a zero-shot setting on the \gp dataset. Lines 1-7 show the original Prompt 0, while Line 8 displays the sentiment predicted by \vicuna. Lines 9-10 are added to the original prompt to request the model to provide probability scores for each class. Lines 11-13 then present the probability scores given by the model. We proceed to extract these probability scores for each class. If the total probability scores do not sum to 1, we normalize the scores by dividing each score by the total sum of the scores.
}

\vspace{8px}
\noindent{\textbf{Fine-tuning sLLMs.}} We fine-tuned all the sLLMs with the training data.
For each epoch, we calculate their macro-F1 score on the validation data.
We fine-tune each sLLM 5 epochs.
We save the best-performing model, i.e., achieving the highest macro-F1 score on the validation data, as the final model.
We then evaluate the best model on the test data. 
We used the following sets of hyper-parameters for all the sLLM: learning rate of 2e-5, batch size of 32, and max length of 256.
\section{Results}
\label{sec:results}

\subsection{RQ1: Impact of different prompts on the performance of bLLMs with zero-shot learning}
It is worth noting that \vicuna and \wizardlm adopt the same style of prompt template; while \llama employs a different prompt template in pre-training, resulting in slight differences in prompt formats\footnote{\url{https://huggingface.co/blog/llama2\#how-to-prompt-llama-2}} (We show an example of \llama prompt in Figure~\ref{fig:few-shot-prompt}, where the zero-shot template is the same template excluding the ``Demonstration'' part).

\begin{table}[t]
\centering
\caption{Zero-shot Performance: Comparative Results of LLMs Across Five Datasets. Cells highlighted in red indicate the highest scores achieved among the three prompts executed by each respective model.}
\label{tab:zero-result}
\small
\begin{tabular}{ccrrrrrr}
\toprule
\multicolumn{1}{l}{} &
\multicolumn{1}{l}{\textit{\textbf{Model}}} &
\multicolumn{1}{l}{\textit{\textbf{Variant}}} &
\multicolumn{1}{l}{\textit{\textbf{Gerrit}}} &
\multicolumn{1}{l}{\textit{\textbf{GitHub}}} &
\multicolumn{1}{l}{\textit{\textbf{GooglePlay}}} &
\multicolumn{1}{l}{\textit{\textbf{Jira}}} &
\multicolumn{1}{l}{\textit{\textbf{StackOverflow}}} \\
\midrule
\multirow{9}{*}{\textbf{Macro-F1}} & &
\textbf{0} &
0.73 &
\cellcolor[HTML]{F4CCCC}0.68 &
\cellcolor[HTML]{F4CCCC}0.89 &
\cellcolor[HTML]{F4CCCC}0.83 &
0.45 \\
&
&
\textbf{1} &
0.71 &
0.64 &
\cellcolor[HTML]{F4CCCC}0.89 &
0.71 &
0.5 \\
&
\multirow{-3}{*}{\textbf{\llama}} &
\textbf{2} &
\cellcolor[HTML]{F4CCCC}0.75 &
\cellcolor[HTML]{F4CCCC}0.68 &
\cellcolor[HTML]{F4CCCC}0.89 &
0.78 &
\cellcolor[HTML]{F4CCCC}0.51 \\
\cmidrule{2-8}
& &
\textbf{0} &
\cellcolor[HTML]{F4CCCC}0.73 &
\cellcolor[HTML]{F4CCCC}0.72 &
\cellcolor[HTML]{F4CCCC}0.98 &
\cellcolor[HTML]{F4CCCC}0.85 &
\cellcolor[HTML]{F4CCCC}0.59 \\
&
&
\textbf{1} &
\cellcolor[HTML]{F4CCCC}0.73 &
0.65 &
0.74 &
0.69 &
0.56 \\
&
\multirow{-3}{*}{\textbf{\vicuna}} &
\textbf{2} &
0.7 &
0.67 &
0.82 &
0.75 &
0.53 \\
\cmidrule{2-8}
&
&
\textbf{0} &
\cellcolor[HTML]{F4CCCC}0.69 &
\cellcolor[HTML]{F4CCCC}0.71 &
0.8 &
0.81 &
0.41 \\
&
&
\textbf{1} &
\cellcolor[HTML]{F4CCCC}0.69 &
0.7 &
\cellcolor[HTML]{F4CCCC}0.82 &
\cellcolor[HTML]{F4CCCC}0.82 &
\cellcolor[HTML]{F4CCCC}0.59 \\
&
\multirow{-3}{*}{\textbf{\wizardlm}} &
\textbf{2} &
0.68 &
0.7 &
0.79 &
0.77 &
0.52 \\
\midrule
\multirow{9}{*}{\textbf{Micro-F1}} &  &
\textbf{0} &
0.82 &
\cellcolor[HTML]{F4CCCC}0.68 &
0.91 &
\cellcolor[HTML]{F4CCCC}0.84 &
0.61 \\
&
&
\textbf{1} &
0.82 &
0.64 &
0.91 &
0.71 &
\cellcolor[HTML]{F4CCCC}0.72 \\
&
\multirow{-3}{*}{\textbf{\llama}} &
\textbf{2} &
\cellcolor[HTML]{F4CCCC}0.83 &
\cellcolor[HTML]{F4CCCC}0.68 &
\cellcolor[HTML]{F4CCCC}0.94 &
0.79 &
0.64 \\
\cmidrule{2-8}
&  &
\textbf{0} &
0.81 &
\cellcolor[HTML]{F4CCCC}0.72 &
\cellcolor[HTML]{F4CCCC}0.97 &
\cellcolor[HTML]{F4CCCC}0.86 &
0.78 \\
&
&
\textbf{1} &
\cellcolor[HTML]{F4CCCC}0.82 &
0.66 &
0.8 &
0.71 &
\cellcolor[HTML]{F4CCCC}0.82 \\
&
\multirow{-3}{*}{\textbf{\vicuna}} &
\textbf{2} &
\cellcolor[HTML]{F4CCCC}0.82 &
0.67 &
0.89 &
0.76 &
0.78 \\
\cmidrule{2-8}
&
&
\textbf{0} &
\cellcolor[HTML]{F4CCCC}0.8 &
\cellcolor[HTML]{F4CCCC}0.71 &
0.86 &
0.82 &
0.65 \\
&
&
\textbf{1} &
\cellcolor[HTML]{F4CCCC}0.8 &
0.7 &
\cellcolor[HTML]{F4CCCC}0.89 &
\cellcolor[HTML]{F4CCCC}0.83 &
\cellcolor[HTML]{F4CCCC}0.73 \\
&
\multirow{-3}{*}{\textbf{\wizardlm}} &
\textbf{2} &
0.79 &
0.7 &
\cellcolor[HTML]{F4CCCC}0.89 &
0.78 &
0.67 \\
\midrule
& & \textbf{0} & \cellcolor[HTML]{F4CCCC}0.65 & \cellcolor[HTML]{F4CCCC}0.64 & \cellcolor[HTML]{F4CCCC}0.88  & 0.80                          & \cellcolor[HTML]{F4CCCC}0.65 \\
&  & \textbf{1} & 0.51                         & 0.59                         & 0.65                          & 0.79                          & 0.63                         \\
& \multirow{-3}{*}{\textbf{\llama}}    & \textbf{2} & 0.49                         & 0.54                         & 0.65                          & \cellcolor[HTML]{F4CCCC}0.81  & 0.47 \\
\cmidrule{2-8}
&  & \textbf{0} & \cellcolor[HTML]{F4CCCC}0.73 & \cellcolor[HTML]{F4CCCC}0.83 & \cellcolor[HTML]{F4CCCC}0.997 & \cellcolor[HTML]{F4CCCC}0.996 & \cellcolor[HTML]{F4CCCC}0.71 \\
& & \textbf{1} & 0.67                         & 0.74                         & 0.96                          & 0.97                          & 0.63  \\
& \multirow{-3}{*}{\textbf{\vicuna}}   & \textbf{2} & 0.54                         & 0.74                         & 0.91                          & 0.95                          & 0.50                         \\
\cmidrule{2-8}
& & \textbf{0} & \cellcolor[HTML]{F4CCCC}0.64 & \cellcolor[HTML]{F4CCCC}0.81 & 0.83                          & \cellcolor[HTML]{F4CCCC}0.97  & 0.59                         \\
& & \textbf{1} & 0.64                         & 0.79                         & \cellcolor[HTML]{F4CCCC}0.95  & 0.96                          & \cellcolor[HTML]{F4CCCC}0.71 \\
\multirow{-9}{*}{\textbf{AUC}} & \multirow{-3}{*}{\textbf{\wizardlm}} & \textbf{2} & 0.57                         & 0.79                         & 0.92                          & 0.96                          & 0.62                        \\
\bottomrule
\end{tabular}
\end{table}

\begin{table}[t]
\centering
\caption{Zero-shot: Score difference between the highest and lowest ones with each LLM on one dataset and the value in the parenthesis shows the difference percentage.}
\label{tab:zero-result-difference}
\small
\begin{tabular}{crrrrrr}
\toprule
\multicolumn{1}{l}{} &
  \textit{\textbf{Model}} &
  \textit{\textbf{Gerrit}} &
  \textit{\textbf{GitHub}} &
  \textit{\textbf{GooglePlay}} &
  \textit{\textbf{Jira}} &
  \textit{\textbf{StackOverflow}} \\
  \midrule
\multirow{3}{*}{\bf Macro-F1} & \textbf{\llama}  & 0.04 (5.6\%) & 0.04 (6.3\%)  & \multicolumn{1}{r}{0} & 0.12 (16.9\%) & 0.06 (13.3\%) \\
\cmidrule{2-7}
& \textbf{\vicuna}   & 0.03 (4.3\%) & 0.07 (10.8\%) & 0.24 (32.4\%)         & 0.16 (23.2\%) & 0.06 (11.3\%) \\
\cmidrule{2-7}
& \textbf{\wizardlm} & 0.01 (1.5\%) & 0.01 (1.4\%)  & 0.03 (3.8\%)          & 0.05 (6.5\%)  & 0.18 (43.9\%) \\
\cmidrule{2-7}
\multicolumn{1}{l}{} &
  \textit{\textbf{Avg. Diff}} &
  \multicolumn{1}{r}{3.80\%} &
  \multicolumn{1}{r}{6.17\%} &
  \multicolumn{1}{r}{12.07\%} &
  \multicolumn{1}{r}{15.53\%} &
  \multicolumn{1}{r}{22.83\%} \\
  \midrule
\multirow{3}{*}{\bf Micro-F1} & \textbf{\llama}    & 0.01 (1.2\%) & 0.04 (6.3\%)  & 0.03 (3.3\%)          & 0.13 (18.3\%) & 0.11 (18.0\%)  \\
\cmidrule{2-7}
& \textbf{\vicuna}   & 0.01 (2.1\%) & 0.06 (12.5\%) & 0.17 (35.4\%)         & 0.15 (31.3\%) & 0.04 (8.3\%)  \\
\cmidrule{2-7}
& \textbf{\wizardlm} & 0.01 (1.3\%) & 0.01 (1.4\%)  & 0.03 (3.5\%)          & 0.05 (6.4\%)  & 0.08 (12.3\%) \\
\cmidrule{2-7}
\multicolumn{1}{l}{} &
  \textit{\textbf{Avg. Diff}} &
  \multicolumn{1}{r}{1.5\%} &
  \multicolumn{1}{r}{6.7\%} &
  \multicolumn{1}{r}{14.0\%} &
  \multicolumn{1}{r}{18.7\%} &
  \multicolumn{1}{r}{12.9\%} \\
  \midrule
  \multirow{3}{*}{\bf AUC} & \textbf{\llama}  & 0.16 (31.9\%) & 0.09 (17.3\%) & 0.23 (36.0\%) & 0.02 (2.3\%) & 0.18 (37.7\%) \\
  \cmidrule{2-7}
  & \textbf{\vicuna}  & 0.19 (34.6\%) & 0.09 (11.8\%) & 0.09 (9.4\%) & 0.05 (5.4\%) & 0.21 (42.1\%)  \\
  \cmidrule{2-7}
  & \textbf{\wizardlm}  & 0.07 (12.8\%) & 0.02 (2.8\%) & 0.13 (15.2\%) & 0.02 (1.7\%) & 0.12 (20.2\%)   \\
  \cmidrule{2-7}
  &  \textit{\textbf{Avg. Diff}} & \multicolumn{1}{r}{26.4\%} & \multicolumn{1}{r}{10.6\%} & \multicolumn{1}{r}{20.2\%} & \multicolumn{1}{r}{3.1\%} & \multicolumn{1}{r}{33.4\%}  \\
  \bottomrule
\end{tabular}
\end{table}

Table~\ref{tab:zero-result} presents the outcomes obtained from our investigation into three distinct bLLMs using three distinct zero-shot prompts. Notably, we observe varying performance levels among these bLLMs when employed with different prompt templates. Furthermore, it is noteworthy that even when using the same bLLM, the optimal prompt template can vary depending on the dataset under consideration.

Specifically, regarding the macro-F1 score, \textit{Prompt 0} emerges as the most effective choice, yielding the highest scores in \textcolor{black}{10} instances. Following closely, \textit{Prompt 1} leads in \textcolor{black}{6} instances.
Interestingly, \textit{Prompt 2}, while achieving the top performance on only \textcolor{black}{4} occasions, occasionally surpasses \textit{Prompt 1} by a significant margin, notably in the case of the \llama within the \sof dataset.
Regarding the micro-F1 scores, both \textit{Prompt 0} and \textit{Prompt 1} achieved the highest scores 7 times, while \textit{Prompt 2} ranked the first 5 times.
\textcolor{black}{Regarding the AUC value, \textit{Prompt 0} emerges as the most effective choice, yielding the highest scores in 12 instances.
\textit{Prompt 1} and \textit{Prompt 2} achieves the top performance on only 2 and 1 occasions, respectively.}
These results show that Prompt 0 can achieve overall best results considering all the metrics.

Table~\ref{tab:zero-result-difference} provides an insight into the variance within each result group under the zero-shot setting.
We define a result \textit{group} as results generated by the same bLLM when applied to the same dataset with varying prompts. This analysis aims to underscore the impact of prompt selection on performance. Remarkably, within the same group, we observe disparities as substantial as 43.9\%. Expanding our examination to encompass different models operating on identical datasets reveals an average difference as substantial as 22.83\%. This discovery underscores the sensitivity of bLLMs to the choice of prompts in zero-shot learning.

\textcolor{black}{
Based on the results of the Wilcoxon signed-rank test with Bonferroni correction for multiple comparisons, we can draw the following conclusions regarding the significance of differences between the distinct prompt templates:
\begin{itemize}
    \item The comparison between prompt 0 and prompt 1 yielded a p-value of 0.0520, which is greater than the corrected alpha threshold of 0.0167. Therefore, we fail to reject the null hypothesis, indicating no significant difference between prompt 0 and prompt 1.
    \item The comparison between prompt 0 and prompt 2 resulted in a p-value of 0.0048, which is less than the corrected alpha threshold of 0.0167. Consequently, we reject the null hypothesis, indicating a significant difference between prompt 0 and prompt 2.
    \item The comparison between prompt 1 and prompt 2 produced a p-value of 1.0, which is much greater than the corrected alpha threshold of 0.0167. Thus, we fail to reject the null hypothesis, indicating no significant difference between prompt 1 and prompt 2.
\end{itemize}
In summary, the analysis reveals a significant difference between prompt 0 and prompt 2, while no significant differences were found between the other pairs of zero-shot prompt templates.
}

\textcolor{black}{
Although the overall difference between different prompt templates on the same model is not significant, the best performance difference between different prompts on the same model can be as large as 43.9\%. Therefore, it is crucial to test multiple prompt templates and select the one that yields the highest accuracy for a given task and dataset. Based on the experimental results, we have the following implications: a) Researchers should test a range of prompts when evaluating models for the SA4SE task, rather than relying on generic prompts. b) Practitioners implementing SA tools for SE domain (e.g., analyzing code reviews or developer forums) should invest time in prompt optimization for their specific use case. c) The variability in results suggests that ensemble methods, combining multiple prompts or models, might offer more robust performance in real-world applications. Our findings also open up an avenue for future research: Developing systematic methods for prompt optimization in SE domain, including but not limited to the SA4SE task.
}

\begin{tcolorbox}[left=4pt,right=4pt,top=2pt,bottom=2pt,boxrule=0.5pt]
\textbf{Answer to RQ1:} 
In the SA4SE context, it is evident that bLLMs exhibit sensitivity to prompts in zero-shot learning scenarios. When employing various prompt templates, the average macro-F1 score difference spans from 3.8\% to 22.83\%, the average micro-F1 score difference ranges from 1.5\% to 18.7\%, \textcolor{black}{and the average AUC value difference varies from 3.1\% to 33.4\%.}
\end{tcolorbox}

\subsection{RQ2: Impact of different shots on the performance of bLLMs with few-shot learning}
\label{subsec:few-shot-learning}

\begin{table}[t]
\centering
\caption{Few-shot Performance: Comparative Results of LLMs Across Five Datasets. Cells highlighted in red indicate the highest scores achieved among the three prompts executed by each respective model.}
\label{tab:few-result}
\small
\begin{tabular}{ccrrrrrr}
\toprule
\multicolumn{1}{l}{} & \multicolumn{1}{l}{\textit{\textbf{Model}}} & \multicolumn{1}{l}{\textit{\textbf{Shot}}} & \multicolumn{1}{l}{\textit{\textbf{Gerrit}}} & \multicolumn{1}{l}{\textit{\textbf{GitHub}}} & \multicolumn{1}{l}{\textit{\textbf{GooglePlay}}} & \multicolumn{1}{l}{\textit{\textbf{Jira}}} & \multicolumn{1}{l}{\textit{\textbf{StackOverflow}}} \\
\midrule
 \multirow{9}{*}{\textbf{Macro-F1}}  &  & \textbf{1} & \cellcolor[HTML]{F4CCCC}0.69 & 0.54 & 0.89 & 0.82 & 0.42 \\
 &  & \textbf{3} & \cellcolor[HTML]{F4CCCC}0.69 & 0.60 & 0.87 & 0.84 & 0.46 \\
& \multirow{-3}{*}{\textbf{\llama}} & \textbf{5} & 0.68 & \cellcolor[HTML]{F4CCCC}0.61 & \cellcolor[HTML]{F4CCCC}1 & \cellcolor[HTML]{F4CCCC}0.89 & \cellcolor[HTML]{F4CCCC}0.47 \\
 \cmidrule{2-8}
 & & \textbf{1} & \cellcolor[HTML]{F4CCCC}0.74 & 0.68 & 0.74 & 0.77 & 0.56 \\
 &  & \textbf{3} & 0.73 & \cellcolor[HTML]{F4CCCC}0.72 & \cellcolor[HTML]{F4CCCC}0.82 & 0.86 & \cellcolor[HTML]{F4CCCC}0.65 \\
 & \multirow{-3}{*}{\textbf{\vicuna}} & \textbf{5} & 0.71 & \cellcolor[HTML]{F4CCCC}0.72 & 0.77 & \cellcolor[HTML]{F4CCCC}0.89 & 0.64 \\
 \cmidrule{2-8}
 &  & \textbf{1} & \cellcolor[HTML]{F4CCCC}0.76 & 0.68 & \cellcolor[HTML]{F4CCCC}0.89 & 0.78 & 0.54 \\
 &  & \textbf{3} & 0.75 & \cellcolor[HTML]{F4CCCC}0.72 & 0.87 & 0.9 & \cellcolor[HTML]{F4CCCC}0.59 \\
 & \multirow{-3}{*}{\textbf{\wizardlm}} & \textbf{5} & 0.75 & 0.71 & 0.82 & \cellcolor[HTML]{F4CCCC}0.91 & 0.54 \\
\midrule
\multirow{9}{*}{\textbf{Micro-F1}} & & \textbf{1} & 0.78 & 0.54 & 0.94 & 0.83 & 0.42 \\
 &  & \textbf{3} & \cellcolor[HTML]{F4CCCC}0.79 & \cellcolor[HTML]{F4CCCC}0.60 & 0.94 & 0.86 & 0.51 \\
 & \multirow{-3}{*}{\textbf{\llama}} & \textbf{5} & 0.77 & \cellcolor[HTML]{F4CCCC}0.60 & \cellcolor[HTML]{F4CCCC}1 & \cellcolor[HTML]{F4CCCC}0.89 & \cellcolor[HTML]{F4CCCC}0.59 \\
\cmidrule{2-8}
 & & \textbf{1} & \cellcolor[HTML]{F4CCCC}0.82 & 0.69 & 0.8 & 0.78 & 0.78 \\
 &  & \textbf{3} & 0.81 & \cellcolor[HTML]{F4CCCC}0.72 & \cellcolor[HTML]{F4CCCC}0.89 & 0.87 & \cellcolor[HTML]{F4CCCC}0.83 \\
 & \multirow{-3}{*}{\textbf{\vicuna}} & \textbf{5} & 0.78 & \cellcolor[HTML]{F4CCCC}0.72 & 0.83 & \cellcolor[HTML]{F4CCCC}0.89 & 0.82 \\
  \cmidrule{2-8}
 &  & \textbf{1} & 0.83 & 0.67 & \cellcolor[HTML]{F4CCCC}0.94 & 0.79 & 0.67 \\
 &  & \textbf{3} & \cellcolor[HTML]{F4CCCC}0.84 & \cellcolor[HTML]{F4CCCC}0.72 & \cellcolor[HTML]{F4CCCC}0.94 & 0.91 & \cellcolor[HTML]{F4CCCC}0.74 \\
 & \multirow{-3}{*}{\textbf{\wizardlm}} & \textbf{5} & 0.83 & 0.70 & 0.91 & \cellcolor[HTML]{F4CCCC}0.92 & 0.74 \\
 \midrule
 & & \textbf{1} & \cellcolor[HTML]{F4CCCC}0.62 & \cellcolor[HTML]{F4CCCC}0.63 & 0.72 & \cellcolor[HTML]{F4CCCC}0.85 & \cellcolor[HTML]{F4CCCC}0.63 \\
 & & \textbf{3} & 0.52                         & 0.62                         & \cellcolor[HTML]{F4CCCC}0.77 & 0.80                         & 0.58                         \\
 & \multirow{-3}{*}{\textbf{\llama}}    & \textbf{5} & 0.57                         & 0.57                         & 0.76                         & 0.83 & 0.58                         \\
 \cmidrule{2-8}
 & & \textbf{1} & 0.71 & \cellcolor[HTML]{F4CCCC}0.75 & \cellcolor[HTML]{F4CCCC}0.96 & 0.92 & \cellcolor[HTML]{F4CCCC}0.65 \\
 & & \textbf{3} & 0.67 & 0.74 & 0.95 & \cellcolor[HTML]{F4CCCC}0.99 & \cellcolor[HTML]{F4CCCC}0.65 \\
 & \multirow{-3}{*}{\textbf{\vicuna}} & \textbf{5} & \cellcolor[HTML]{F4CCCC}0.73 & \cellcolor[HTML]{F4CCCC}0.75 & \cellcolor[HTML]{F4CCCC}0.96 & 0.98 & \cellcolor[HTML]{F4CCCC}0.65 \\
 \cmidrule{2-8}
 &  & \textbf{1} & \cellcolor[HTML]{F4CCCC}0.74 & \cellcolor[HTML]{F4CCCC}0.72 & 0.92 & 0.92 & 0.56                         \\
 &  & \textbf{3} & 0.63  & \cellcolor[HTML]{F4CCCC}0.72 & \cellcolor[HTML]{F4CCCC}0.93 & \cellcolor[HTML]{F4CCCC}0.99 & \cellcolor[HTML]{F4CCCC} 0.58 \\
\multirow{-9}{*}{\textbf{AUC}} & \multirow{-3}{*}{\textbf{\wizardlm}} & \textbf{5} & 0.59  & 0.71 & 0.87 & 0.98 & 0.53 \\
\bottomrule
\end{tabular}
\end{table}

Table~\ref{tab:few-result} showcases the outcomes of few-shot learning utilizing three distinct bLLMs across five diverse datasets. It is important to reiterate that a result ``group'' signifies the results produced by the same bLLM when applied to the same dataset with varying shot numbers.

In summary, when considering macro-F1, the 5-shot configuration emerges as the leader in 7 instances, followed by the 3-shot setup in 5 cases, and the 1-shot configuration excels in 4 instances. 
Regarding micro-F1, the 5-shot configuration is at the top 7 times, the 3-shot setup prevails 9 times, and the 1-shot configuration leads twice. 
Generally, the trend for both macro-F1 and micro-F1 indicates that having more than one example is more beneficial than having only one.

Among the three models, \llama consistently benefits from having more examples, except for the \gr dataset concerning macro-F1 and micro-F1. In the case of \vicuna and \wizardlm, the impact of additional examples is noticeable primarily on the \jr dataset, affecting both macro-F1 and micro-F1. This underscores the fact that the influence of additional examples can vary depending on the bLLM and dataset employed. This can be explained by recognizing that SA, especially the sentiment classification task central to our work, is relatively straightforward. Consequently, the effects of increasing training examples might not be as pronounced.
\textcolor{black}{All bLLMs show improved performance with more shots on the \jr dataset, likely due to its binary classification nature. Despite being prompted to output only [\textit{``negative''}, \textit{``positive''}] labels, bLLMs tend to predict \textit{neutral} sentiments. This tendency decreases as the number of shots in the prompt increases. With only two labels in the \jr dataset, fewer ``neutral'' predictions lead to higher macro- and micro-F1 scores. This suggests that additional shots significantly reduce bLLMs' misclassification tendencies, especially in datasets with limited label options like \jr.
}

\textcolor{black}{We take \vicuna's performance as an example to illustrate this trend. In zero-shot settings, it explicitly predicts ``neutral'' 19 times on the \jr dataset with the Prompt 1.  This issue diminishes in few-shot settings: \vicuna predicts ``neutral'' 11, 8, and 5 times with 1-shot, 3-shot, and 5-shot prompt templates, respectively.}

\textcolor{black}{Moving on from the advantages of more shots, we also observed a decline in macro-F1 scores with an increase in the number of examples. This trend is particularly noticeable when} applying all three bLLMs to the \gr dataset and when using \wizardlm on the \gp dataset. 
One plausible explanation for this phenomenon is that an increased number of examples leads to longer prompts, which could potentially confuse the bLLMs.
As indicated by Table~\ref{tab:dataset}, the documents from the \gr dataset have the highest average number of tokens. Consequently, introducing more examples results in even lengthier prompts. This observation aligns with prior research on bLLMs in the broader SA domain~\cite{zhang2023sentiment}.

\begin{figure}[t]
\begin{subfigure}{0.34\textwidth}
        \centering
        \includegraphics[width=\linewidth]{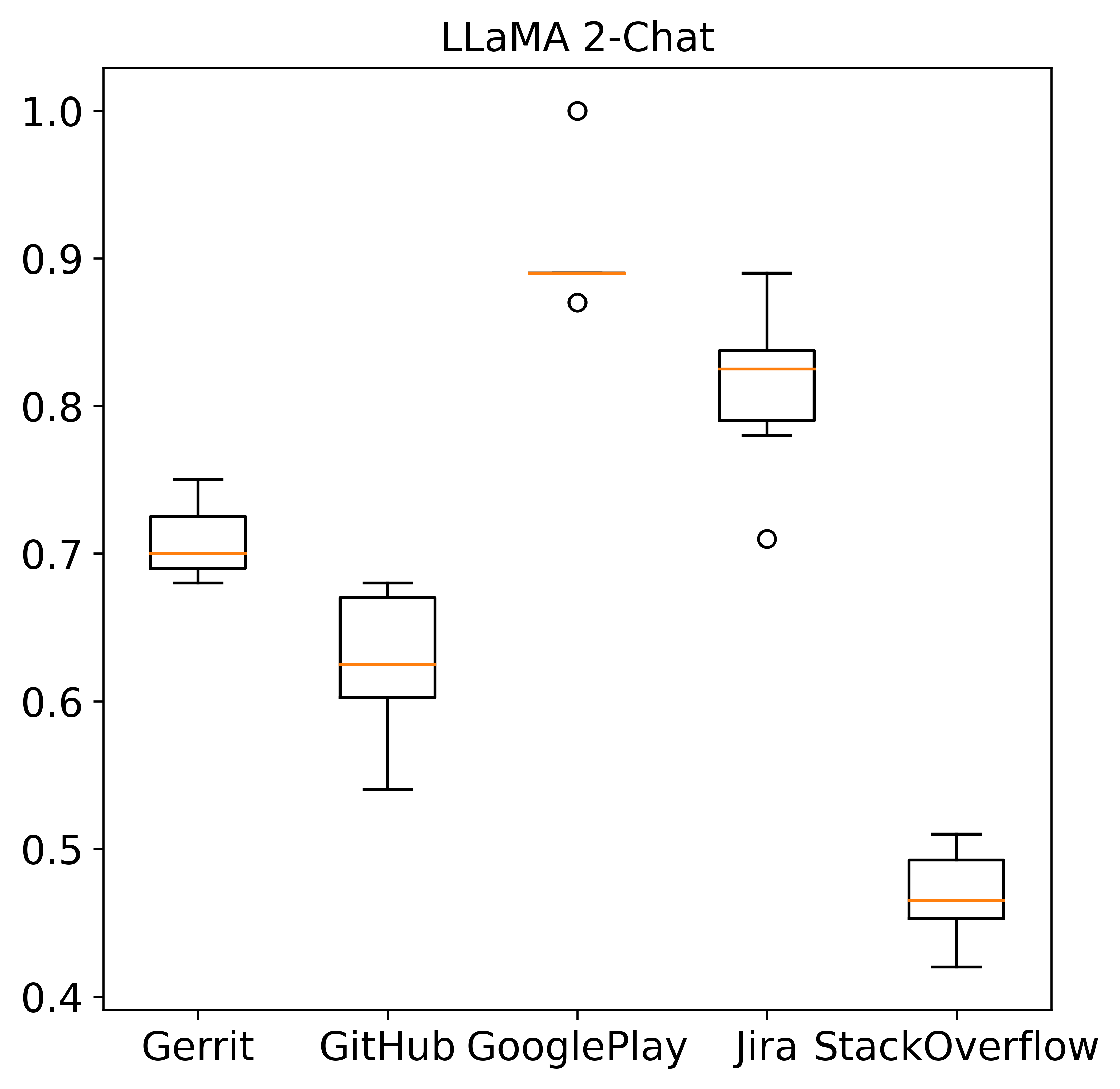}
    \end{subfigure}%
    \begin{subfigure}{0.34\textwidth}
        \centering
        \includegraphics[width=\linewidth]{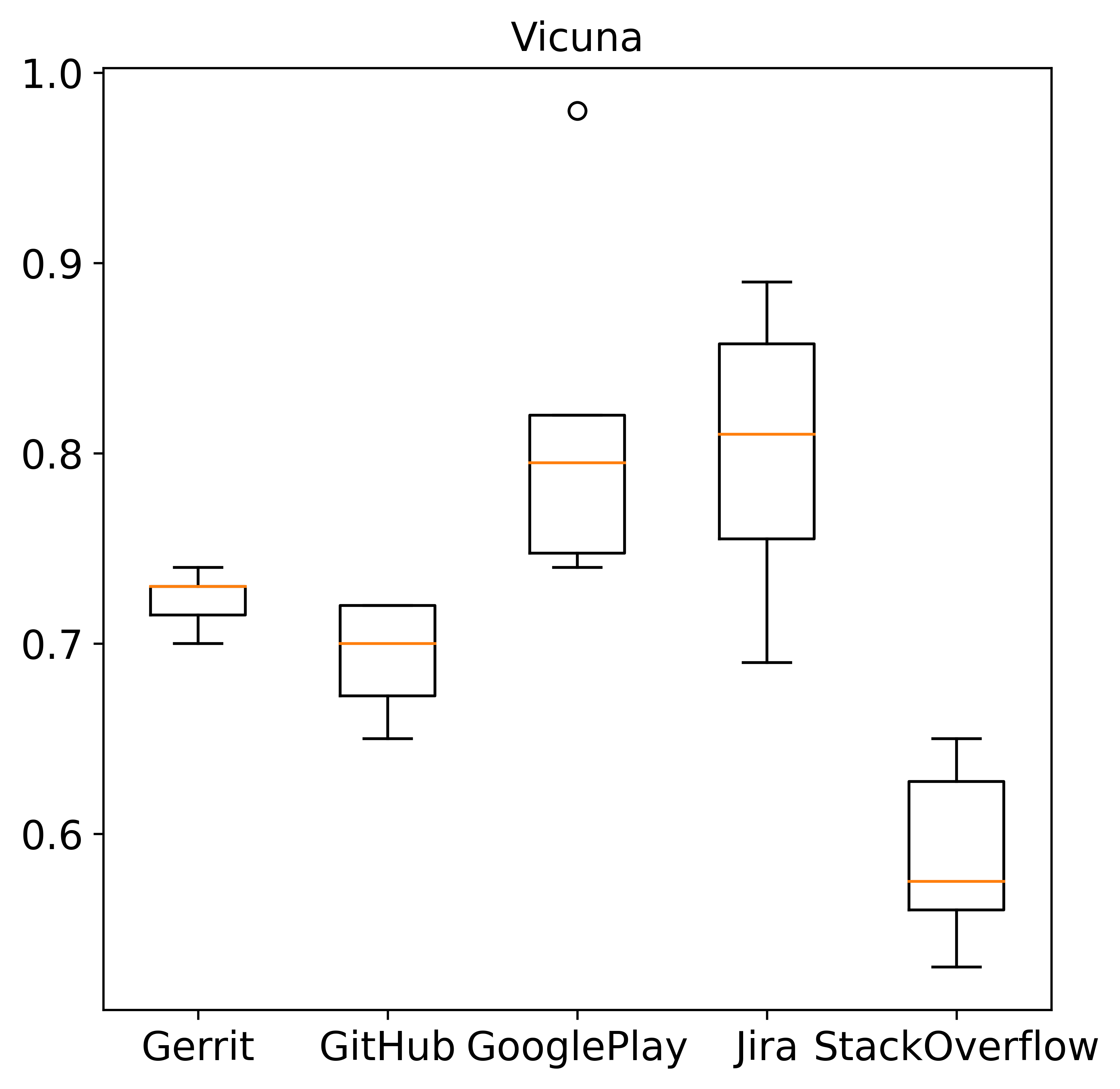}
    \end{subfigure}%
    \begin{subfigure}{0.34\textwidth}
        \centering
        \includegraphics[width=\linewidth]{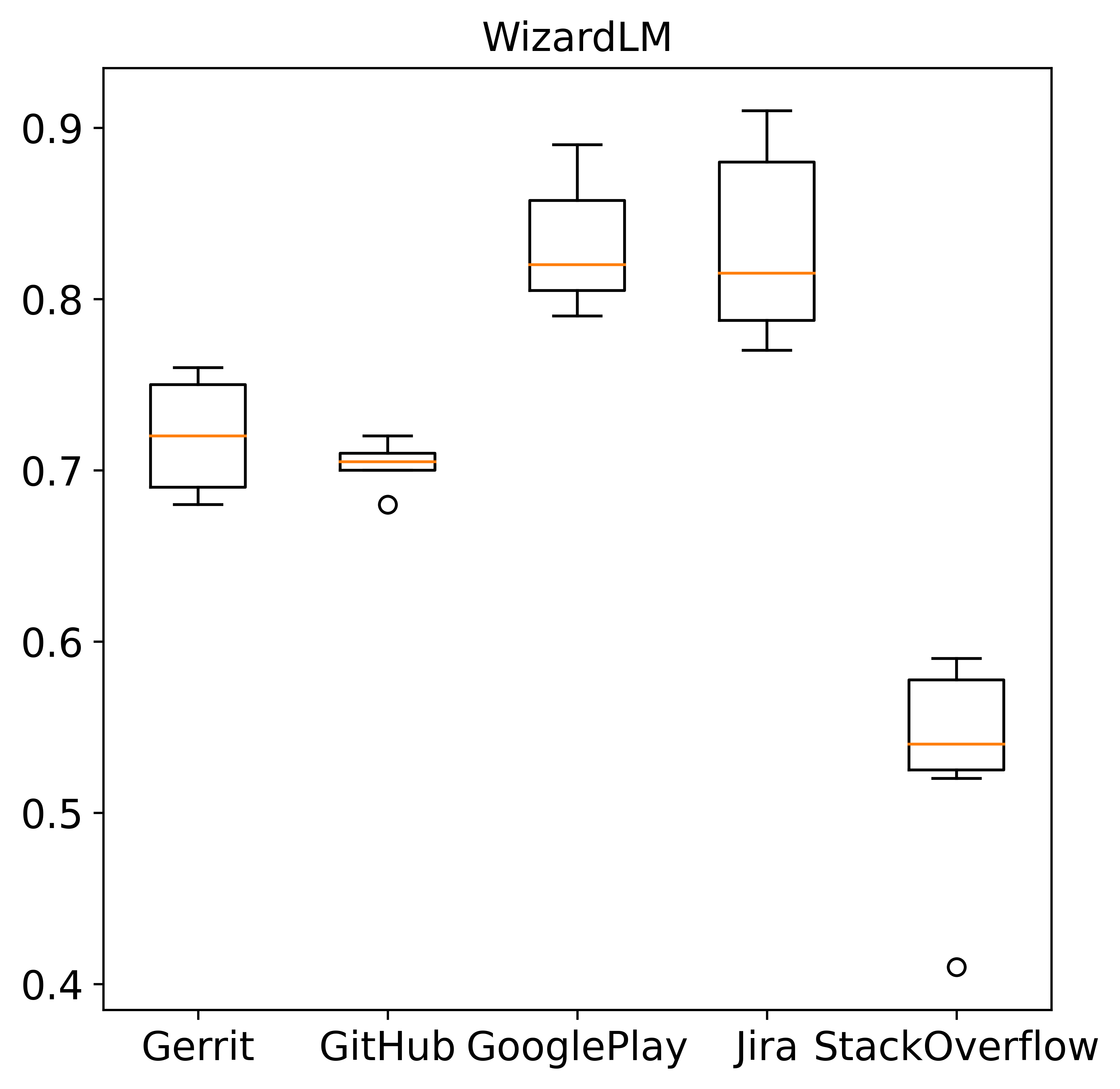}
    \end{subfigure}
    \caption{Sensitivity of different prompt designs. The circles depicted in the figure represent outlier data points.}
    \label{fig:boxplot}
\end{figure}

Figure~\ref{fig:boxplot} provides a clearer illustration of this phenomenon. The box plot delves into the variance of macro-F1 scores achieved by different prompts for each model on each dataset. 
\textcolor{black}{We have examined our study's six prompt templates.}
This figure reveals that the influence of different prompts on performance varies depending on both the model and the dataset.
In general, the models demonstrate differing levels of sensitivity to prompts.
Notably, on the \jr dataset, all models exhibit high sensitivity to prompts, signifying that the choice of prompt has a substantial impact on results. In contrast, on the \gr and \gh datasets, models appear less responsive to different prompts, suggesting that the choice of prompt has a relatively smaller effect on their performance.

\begin{figure}[t]
    \begin{subfigure}{0.5\textwidth}
        \centering
        \includegraphics[width=\linewidth]{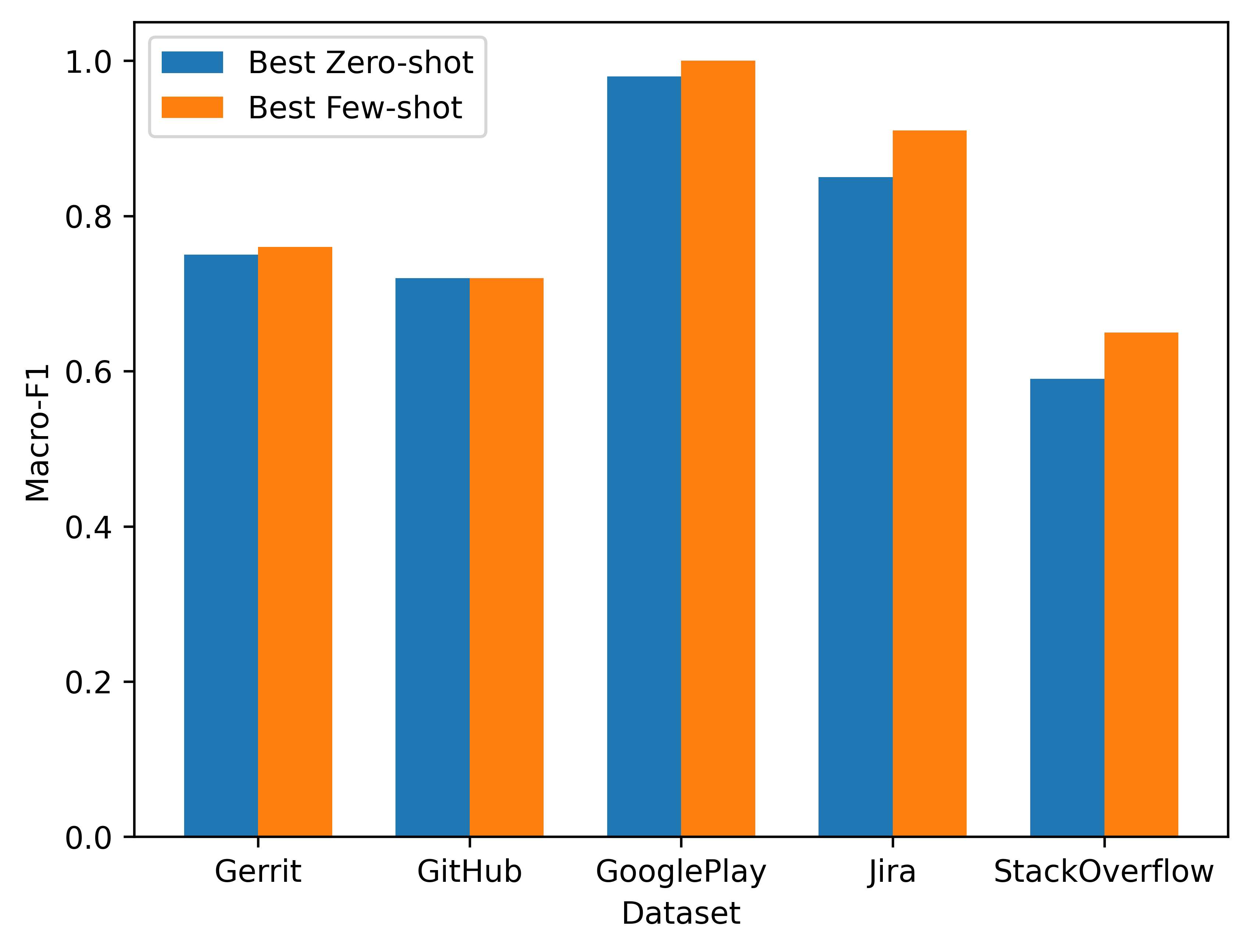}
    \end{subfigure}%
    \begin{subfigure}{0.5\textwidth}
        \centering
        \includegraphics[width=\linewidth]{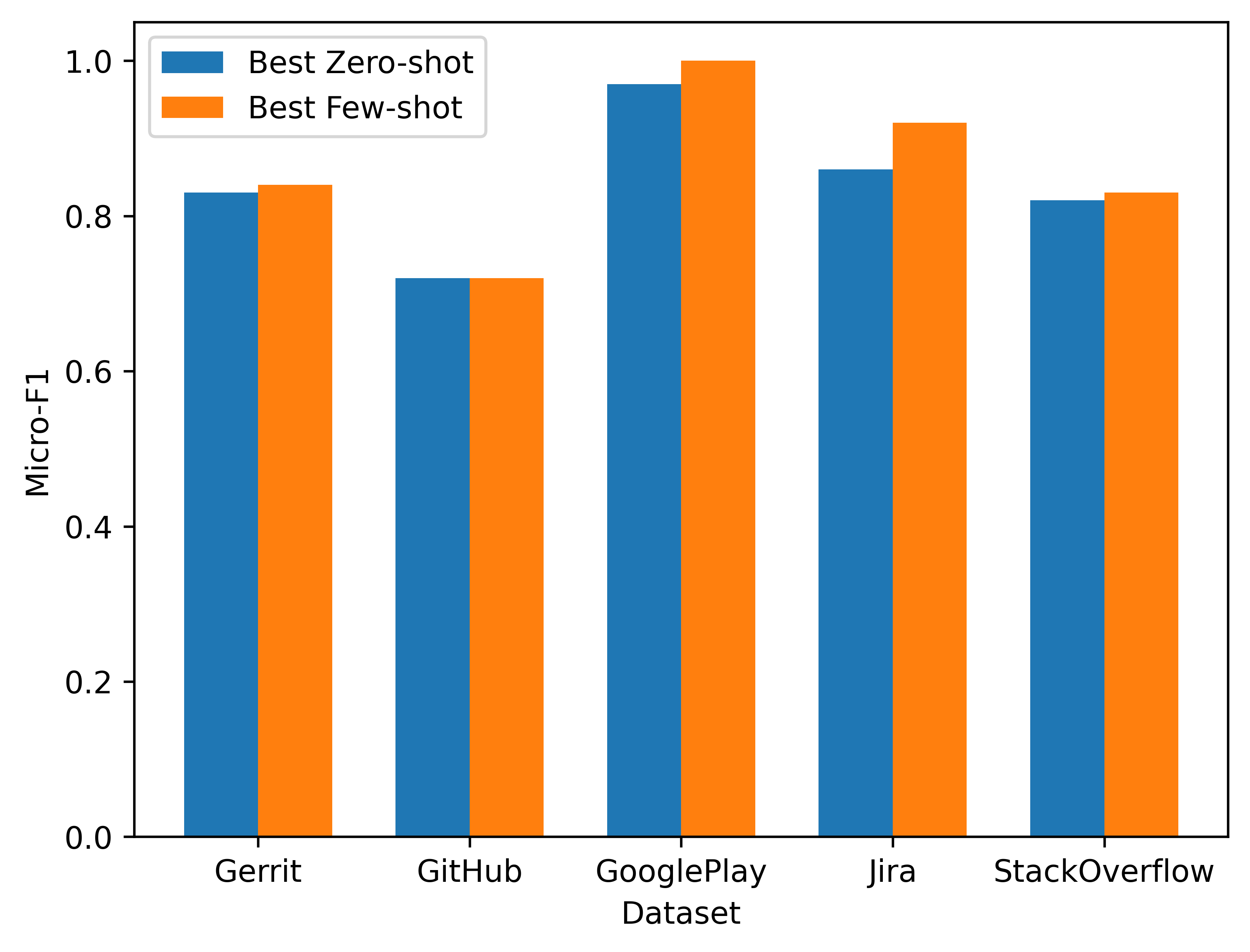}
    \end{subfigure}
    \caption{Comparison of the highest macro-F1 and micro-F1 scores achieved through zero-shot learning and few-shot learning.}
    \label{fig:zsl_fsl}
\end{figure}

We also compare the best results of any of the three bLLMs under few-shot learning and those achieved under zero-shot learning.
Figure~\ref{fig:zsl_fsl} illustrates that, in all the five datasets, the highest macro-F1 score achieved through few-shot learning either surpasses or equals the highest macro-F1 score attained via zero-shot learning. 
We can observe a similar trend in terms of the micro-F1 scores.
This trend is particularly pronounced on the \jr dataset, where bLLMs perform notably better under the few-shot learning paradigm.
\textcolor{black}{Furthermore, to assess whether the results of few-shot learning are significantly better than zero-shot learning, we conducted a Wilcoxon signed-rank test on one pairwise comparison: the zero-shot results from \textit{Prompt 0} versus the five-shot results. Notably, this comparison yielded a p-value of 0.674. This leads us to retain the null hypothesis (H0.2), indicating that the observed differences in results did not reach statistical significance.}
Hence, it is important to recognize that while few-shot learning does demonstrate superior results, the margin of improvement is not substantial.

\begin{tcolorbox}[left=4pt,right=4pt,top=2pt,bottom=2pt,boxrule=0.5pt]
\textbf{Answer to RQ2:} 
Although the top-performing bLLM achieved equal or higher macro- and micro-F1 scores in all five datasets with few-shot learning, \textcolor{black}{the performance gap between few-shot and zero-shot learning} was insignificant. In addition, there is no guarantee that the same bLLM will exhibit improved performance through few-shot learning over zero-shot learning.
\end{tcolorbox}

\subsection{RQ3: Comparison between fine-tuned sLLMs and bLLMs}
\begin{table}[t]
\centering
\caption{Results of LLMs compared with fine-tuned sLLMs. Cells highlighted in red indicate the highest scores achieved among the three prompts executed by each respective model.}
\label{tab:comparison-result}
\small
\begin{tabular}{cccrrrrr}
\toprule
\multicolumn{1}{l}{} 
&  & 
\textit{\textbf{Model}} & \multicolumn{1}{l}{\textit{\textbf{Gerrit}}} & \multicolumn{1}{l}{\textit{\textbf{GitHub}}} & \multicolumn{1}{l}{\textit{\textbf{GooglePlay}}} & \multicolumn{1}{l}{\textit{\textbf{Jira}}} & \multicolumn{1}{l}{\textit{\textbf{StackOverflow}}} \\
\midrule
 &  & \textbf{\llama} & 0.75 & 0.68 & \cellcolor[HTML]{F4CCCC}1 & 0.89 & 0.51 \\
 &  & \textbf{\vicuna} & 0.74 & 0.72 & 0.98 & 0.89 & 0.65 \\
 & \multirow{-3}{*}{\textbf{bLLM}} & \textbf{\wizardlm} & 0.76 & 0.72 & 0.89 & 0.91 & 0.59 \\
 \cmidrule{2-8}
 & & \textbf{ALBERT} & 0.73 & 0.90 & 0.56 & \cellcolor[HTML]{F4CCCC}0.97 & 0.64 \\
 & & \textbf{BERT} & 0.76 & 0.92 & 0.49 & 0.95 & 0.50 \\
 & & \textbf{DistilBERT} & \cellcolor[HTML]{F4CCCC}0.81 & 0.92 & 0.57 & 0.95 & 0.60 \\
 & & \textbf{RoBERTa} & 0.74 & \cellcolor[HTML]{F4CCCC}0.94 & 0.42 & 0.95 & \cellcolor[HTML]{F4CCCC}0.68 \\
\multirow{-8}{*}{\textbf{Macro-F1}} & \multirow{-5}{*}{\textbf{sLLM}} &\textbf{XLNet} & 0.77 & 0.91 & 0.39 & 0.94 & 0.67 \\
\midrule
 &  & \textbf{\llama} & 0.83 & 0.68 & \cellcolor[HTML]{F4CCCC}1 & 0.89 & 0.72 \\
 & & \textbf{\vicuna} & 0.82 & 0.72 & 0.97 & 0.89 & 0.83 \\
 & \multirow{-3}{*}{\textbf{bLLM}} & \textbf{\wizardlm} & 0.84 & 0.72 & 0.94 & 0.92 & 0.74 \\
 \cmidrule{2-8}
 & & \textbf{ALBERT} & 0.81 & 0.90 & 0.80 & \cellcolor[HTML]{F4CCCC}0.97 & 0.84 \\
 & & \textbf{BERT} & 0.80 & 0.92 & 0.71 & 0.96 & 0.84 \\
 & & \textbf{DistilBERT} & \cellcolor[HTML]{F4CCCC}0.86 & 0.92 & 0.83 & 0.96 & 0.84 \\
 & & \textbf{RoBERTa} & 0.81 & \cellcolor[HTML]{F4CCCC}0.94 & 0.63 & 0.95 & 0.86 \\
\multirow{-8}{*}{\textbf{Micro-F1}} & \multirow{-5}{*}{\textbf{sLLM}} & \textbf{XLNet} & 0.83 & 0.91 & 0.57 & 0.95 & \cellcolor[HTML]{F4CCCC}0.89 \\
\midrule
&  & \textbf{\llama} & 0.65 & 0.64 & 0.88 & 0.85 & 0.65 \\
 &  & \textbf{\vicuna} & 0.73 & 0.83 & \cellcolor[HTML]{F4CCCC}0.997 & 0.996 & 0.71 \\
 & \multirow{-3}{*}{\textbf{bLLM}} & \textbf{\wizardlm} & 0.74 & 0.81 & 0.95 & 0.99 & 0.71 \\
 \cmidrule{2-8}
 &  & \textbf{ALBERT} & 0.83 & 0.98 & 0.67 & \cellcolor[HTML]{F4CCCC}0.998 & \cellcolor[HTML]{F4CCCC}0.87 \\
 &  & \textbf{BERT} & 0.89 & 0.98 & 0.80 & \cellcolor[HTML]{F4CCCC}0.998 & 0.81 \\
 &  & \textbf{DistilBERT} & \cellcolor[HTML]{F4CCCC}0.89 & \cellcolor[HTML]{F4CCCC}0.99 & 0.89 & 0.996 & 0.82 \\
 &  & \textbf{RoBERTa} & 0.87 & \cellcolor[HTML]{F4CCCC}0.99 & 0.69 & 0.99 & \cellcolor[HTML]{F4CCCC}0.87 \\
\multirow{-8}{*}{\textbf{AUC}} & \multirow{-5}{*}{\textbf{sLLM}} & \textbf{XLNet} & \cellcolor[HTML]{F4CCCC}0.89 & 0.98 & 0.73 & 0.99 & 0.86 \\
\bottomrule
\end{tabular}
\end{table}

Table~\ref{tab:comparison-result} presents a comparative analysis of the top-performing results obtained through two distinct approaches: prompting bLLMs and fine-tuning sLLMs.

\vspace{4px}
\noindent{On the \gp dataset, where the fine-tuning data is limited to fewer than 300 data points and with a negative:neutral:positive ratio of 26:5:37, bLLMs demonstrate a remarkable performance advantage. The most effective bLLM, \llama, significantly enhances the performance of the leading sLLM, \db, by a substantial improvement of 75.4\%.
It indicates that when labeled data is scarce and the training dataset is highly imbalanced, fine-tuning bLLMs should be the preferred approach over sLLMs.}

\vspace{4px}
\noindent{On the contrary, we have observed that fine-tuning sLLMs produces superior results on the \gh and \jr datasets. The \gh dataset benefits from a larger training dataset and a more evenly distributed class structure, making it particularly well-suited for fine-tuning sLLMs. RoBERTa outperforms the best-performing bLLMs \vicuna and \wizardlm by 30.6\%.}
In contrast, the \jr dataset, though smaller than \gh, offers a more favorable class distribution than \gp, with a negative-to-positive ratio of approximately 1:2.
The best-performing sLLM, ALBERT, outperforms the best-performing bLLM, \wizardlm by 6.6\%.

\vspace{4px}
\noindent{For the \gr and \sof dataset, both bLLMs and sLLMs exhibit relatively modest performance. 
The best-performing bLLM achieves a macro-F1 score of 0.65, while the top sLLM achieves a macro-F1 score of 0.68 on the \sof dataset. 
The corresponding score is 0.76 and 0.81 on the \gr dataset.
Both datasets share the challenge of imbalanced label distribution, which explains the limited success of sLLMs. 
Furthermore, for the \sof dataset, given the short length of the sentences in this dataset, neither bLLMs nor sLLMs may have sufficient contextual information to make accurate sentiment predictions.
}

\begin{figure}[t]
    \begin{subfigure}{0.5\textwidth}
        \centering
        \includegraphics[width=\linewidth]{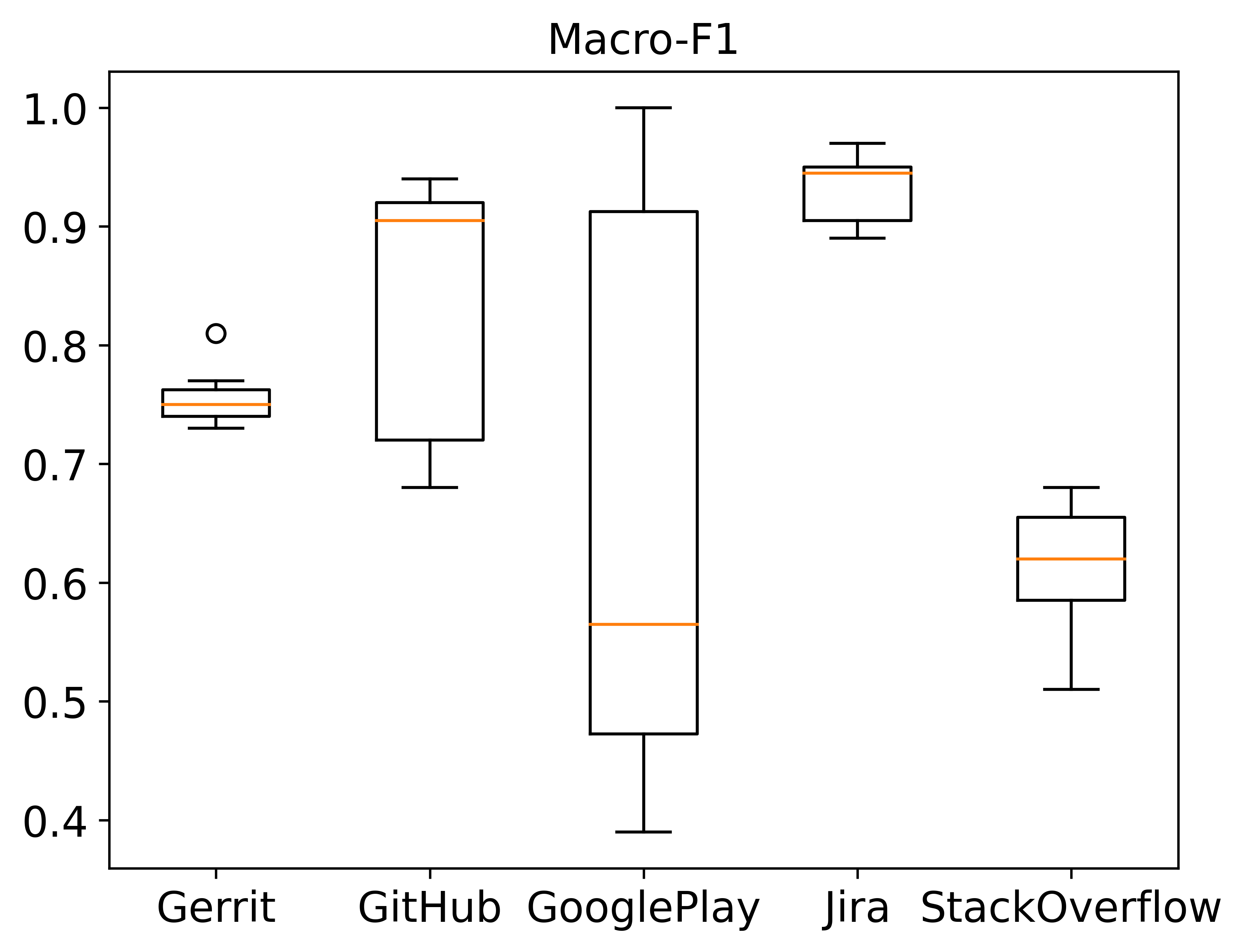}
    \end{subfigure}%
    \begin{subfigure}{0.5\textwidth}
        \centering
        \includegraphics[width=\linewidth]{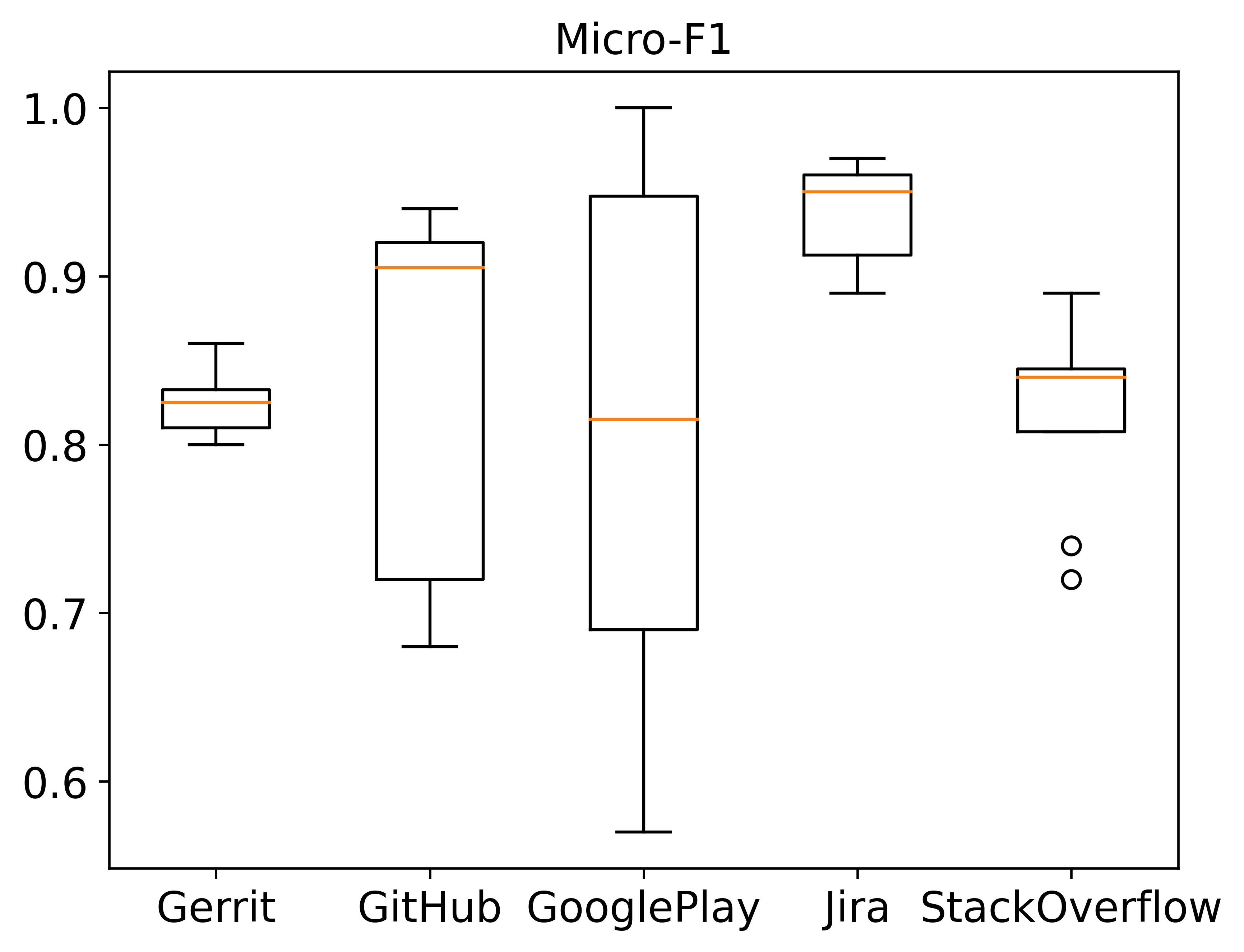}
    \end{subfigure}
    \caption{Performance variance of all the models on each dataset.}
    \label{fig:all_box_plot}
\end{figure}

\vspace{4px}
Moreover, in Figure~\ref{fig:all_box_plot}, we observe the performance of all the models across different datasets. Notably, on the \gp dataset, there is a significant variance, with bLLMs standing out by achieving the highest macro-F1 and micro-F1 scores. This underscores bLLMs' superiority over sLLMs on this specific dataset.

On the other hand, for the \gr, \jr, and \sof datasets, the variance is comparatively smaller, suggesting that while sLLMs outperform bLLMs, the margin of difference is not substantial. 

Conversely, in the case of the \gh dataset, sLLMs demonstrate a substantial advantage over bLLMs. This discrepancy is likely due to the abundance of training data available for the \gh dataset. It reinforces that sLLMs are most effective when ample, balanced training data is available.

\textcolor{black}{To evaluate the significance of differences between bLLMs and sLLMs, we performed a Wilcoxon signed-rank test on one pair of comparisons: the overall best bLLM, i.e., \llama, versus the overall best sLLM, i.e., RoBERTa. The comparison resulted in a p-value of 0.277, leading us to retain the null hypothesis (H0.3). This indicates that the observed differences in results did not achieve statistical significance.}

\begin{tcolorbox}[left=4pt,right=4pt,top=2pt,bottom=2pt,boxrule=0.5pt]
\textbf{Answer to RQ3:} 
In scenarios with limited labeled data and pronounced class imbalance, prompting bLLMs is a more effective strategy, outperforming fine-tuning sLLMs, e.g., in the \gp dataset, \llama outperforms the \db by 75.4\%. In contrast, when ample training data is available and the dataset demonstrates a balanced distribution, the preference should lean toward sLLMs as the more suitable approach, e.g., in the \gh dataset, RoBERTa outperforms \vicuna by 30.6\%.
\end{tcolorbox}

\subsection{RQ4: Error Analysis}
\label{sec:error}

\begin{figure}
    \begin{subfigure}{0.5\textwidth}
        \centering
        \includegraphics[width=0.8\linewidth]{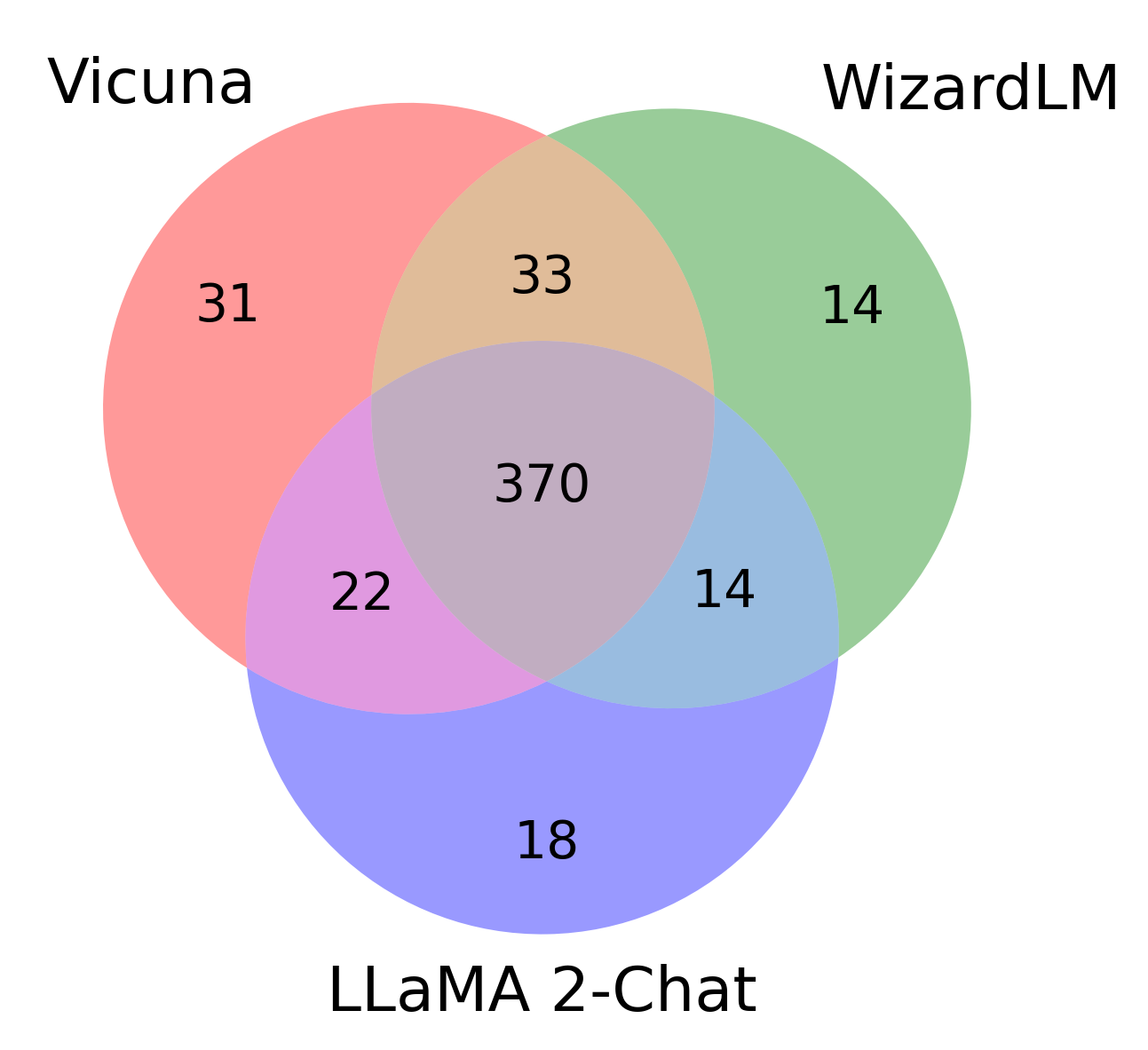}
        \caption{Zero-shot learning with Prompt 0}
    \end{subfigure}%
    \begin{subfigure}{0.5\textwidth}
        \centering
        \includegraphics[width=0.8\linewidth]{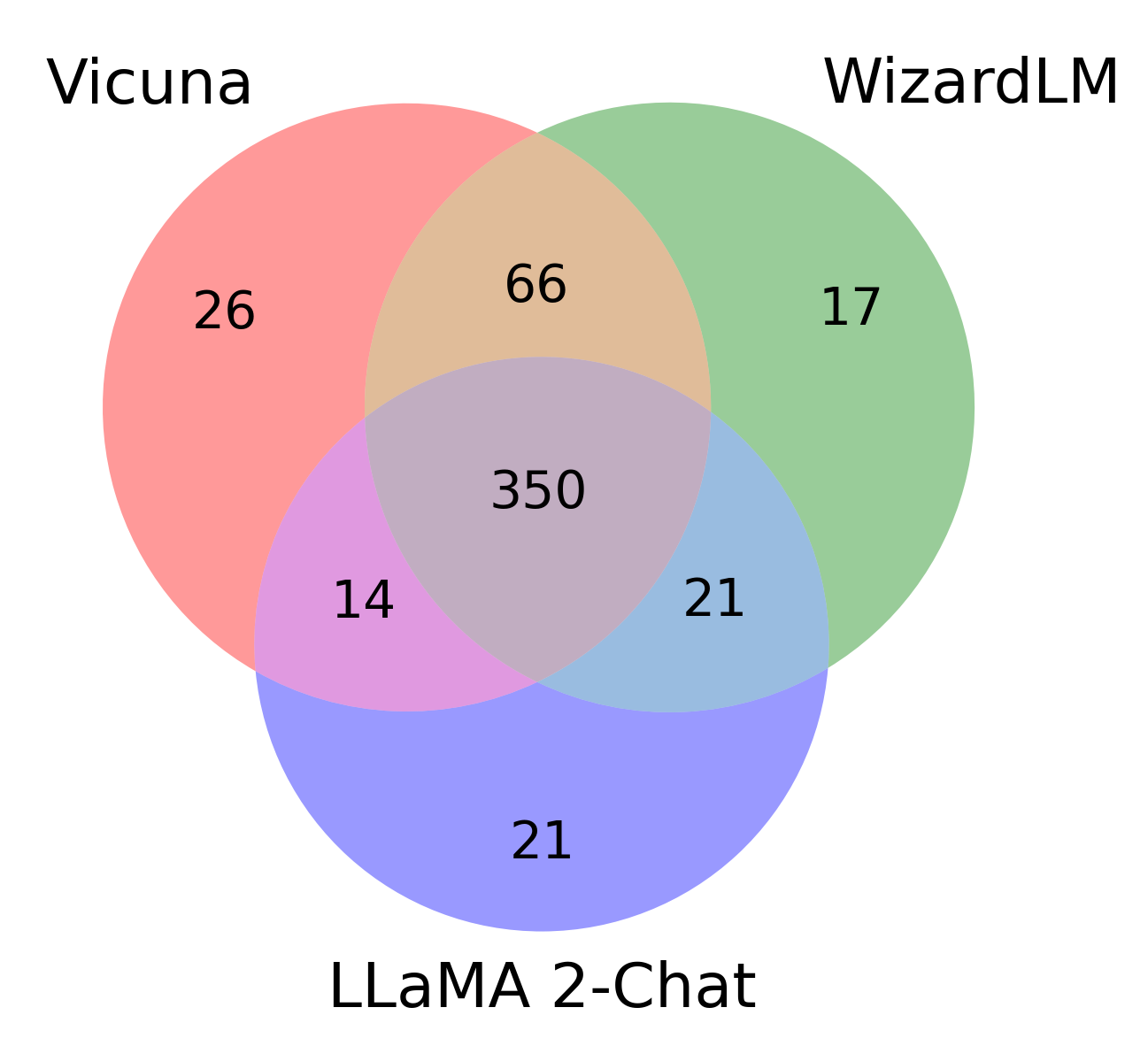}
        \caption{Few-shot learning with 5-shot prompt}
    \end{subfigure}
    \caption{The Venn diagram of the correct predictions made by bLLMs.}
    \label{fig:venn}
\end{figure}

\textbf{Quantitative Analysis.}
In Figure~\ref{fig:venn}, it is evident that all three bLLMs consistently generate accurate predictions in a substantial portion of instances. Specifically, in the realm of zero-shot learning, these models collectively predicted the correct sentiments in 73.7\% cases, and in the context of few-shot learning, they made the correct prediction in 61.4\% cases. 
Notably, both \vicuna and \wizardlm stand out as the bLLMs with the highest degree of overlapping predictions across both scenarios. In the zero-shot context, they share common predictions in an impressive 83.3\% of their respective correct predictions, while in the few-shot scenario, this shared correct prediction rate remains substantial at 70.1\%. Conversely, the lowest degree of common correct predictions is observed between \vicuna and \llama, with rates of 80.3\% and 69.1\% for zero-shot and few-shot scenarios, respectively. These results underscore the bLLMs' capacity to achieve comparable success rates in most cases while also emphasizing their unique strengths.

\begin{table}[t]
\centering
\caption{Overlap in Misclassification Across LLMs in Zero-Shot and Few-Shot Settings. The \textit{Common} Column Indicates Misclassifications Shared Between Both Settings.}
\label{tab:error_prediction}
\begin{tabular}{llllr}
\toprule
 & \multicolumn{3}{c}{\textbf{Misclassified (\% of the test set)}} & \multicolumn{1}{c}{\multirow{2}{*}{\textit{\textbf{Test set size}}}} \\
 \cmidrule{2-4}
 & \multicolumn{1}{c}{\textit{\textbf{Zero-shot}}} & \multicolumn{1}{c}{\textit{\textbf{Few-shot}}} & \textit{\textbf{Common}} & \multicolumn{1}{c}{} \\
 \midrule
\textit{\textbf{Gerrit}} & 15 (13.2\%) & 10 (8.8\%) & 8 (7.0\%) & 114 \\
\midrule
\textit{\textbf{GitHub}} & 39 (15.6\%) & 41 (16.4\%) & 23 (9.2\%) & 250 \\
\midrule
\textit{\textbf{GooglePlay}} & 1 (2.9\%) & 0 & 0 & 35 \\
\midrule
\textit{\textbf{Jira}} & 8 (10.5\%) & 4 (5.3\%) & 4 (5.3\%) & 76 \\
\midrule
\textit{\textbf{StackOverflow}} & 19 (17.4\%) & 14 (12.8\%) & 11 (10.1\%) & 109 \\
\midrule
\midrule
\textit{\textbf{Total}} & 82 (14\%) & 69 (11.8\%) & 46 (7.9\%) & 584 \\
\bottomrule
\end{tabular}
\end{table}

Now, we shift our focus to the errors made by these bLLMs.
Table~\ref{tab:error_prediction} demonstrates again that overall, few-shot learning is more effective than zero-shot learning, as all the bLLMs misclassified more cases under zero-shot learning.
However, we also notice that the number of common misclassification by both settings is 46, which accounts for 7.9\% of the total test cases.
To better understand the difficulties and challenges faced by bLLMs on the task of SA4SE, we manually examined these cases.
In our experiments, we analyze the cases where all the bLLMs did not get any correct prediction no matter the setting.
\vspace{8px}

\begin{table}[t]
\centering
\caption{Distribution of error categories and their percentage among the whole error cases.}
\label{tab:error_category}
\begin{tabular}{llllr}
\toprule
\textbf{Error Category} & \textbf{\# Cases (\%)} \\
\midrule
Polar facts & 16 (34.8\%) \\
Subjectivity in annotation & 13 (28.3\%) \\
General error & 8 (17.4\%) \\
Politeness & 5 (10.9\%) \\
Implicit sentiment polarity & 4 (8.7\%) \\
\bottomrule
\end{tabular}
\end{table}
\noindent{\textbf{Qualitative Analysis.}}
Table~\ref{tab:error_category} shows the error categories of these common misclassifications by all the bLLMs. 
\textit{Polar facts} emerge as the most prominent category, representing the majority of failure cases.
In some cases, the sentence describes a fact, which may usually invoke for most people a positive or negative feeling, i.e. the annotator considered the described situation either as desirable or undesirable.
They have been annotated inconsistently across different datasets.
For example, in \gh and \sof datasets, they were labeled as \textit{neutral}. 
However, in \gr and \jr datasets, they were labeled as \textit{negative}.
For instance, in the \jr dataset, \textit{There is no need to reference \$wizard, it's an object} was labeled as negative.
In the \gh dataset, \textit{Ok, I'll fix them.} was labeled as neutral.

The second most prevalent category, \textit{Subjectivity in annotation}, comprises cases where the evaluators' interpretation of sentiment differed from the originally assigned label.
As recognized in prior works~\cite{liu2010sentiment}, sentiment or option itself is subjective.
Similarly, sentiment annotation is also a subjective activity.
Depending on personality traits or personal disposition, different annotators' perceptions of emotions might vary~\cite{scherer2004emotions}.
Thus, it is not rare that the evaluators have different perceptions from the original annotators. One instance is from the \sof dataset, \textit{Less likely to be blocked by paranoid firewall configurations.}. The evaluators consider this sentence as \textit{positive}, while the ground-truth label is \textit{negative}. Depending on which perspective we think, both make sense: we consider it positive, as we focus on ``less likely to be blocked.'' However, the annotators may give more weight to ``paranoid firewall configurations''.

\textit{General errors} account for 17.4\% of cases and occur when the model fails to identify clues in the document that would be readily apparent to a human. For instance, emoticons can signal sentiment, as observed in the sentence from the \gh dataset, ``yep, it's work, but I need to add user and password for proxy connection=('.'' This sentence may convey negative sentiment to a human, particularly due to the emoticon ``=('' embedded within it.

\textit{Politeness} contributes to 10.9\% of the error cases, arising when the presence of phrases like ``thanks'' or ``sorry'' leads to inconsistencies across different datasets. For instance, in the \gh dataset, the sentence ``sorry I did not realize you were already there...'' was labeled as ``negative'', although some individuals may perceive it as ``neutral''. Similarly, in the \sof dataset, ``Good luck!'' was labeled as ``neutral'', but certain interpretations could classify it as ``positive''. These inconsistencies pose challenges for models when predicting labels with limited examples.

Lastly, \textit{Implicit sentiment polarity} accounts for 8.7\% error cases. When there is a lack of explicit sentiment clues, it could be hard to decide which sentiment is contained. For instance, \textit{Yes, it did not cause message loss just unnecessary retransmits.}, this sentence was annotated as negative. However, there is no obvious sentiment clue.

In summary, due to the inconsistency in labeling rules and the subjective nature of the task, challenges arise where bLLMs may struggle to improve significantly. However, in the case of general errors, there is a potential for improvement as bLLMs continue to advance.

\begin{tcolorbox}[left=4pt,right=4pt,top=2pt,bottom=2pt,boxrule=0.5pt]
\textbf{Answer to RQ4:} 
In 7.9\% of test cases, all the bLLMs consistently fail to make correct predictions, regardless of the setting. 
The inconsistency in labeling rules and the inherently subjective nature of SA4SE remain challenges for bLLMs.
\end{tcolorbox}
\section{Discussion}
\label{sec:discussion}

\begin{table}[!htbp]
\caption{Comparative results of best fine-tuned sLLMs and traditional machine learning models using sLLMs embeddings. Cells highlighted in red indicate the highest scores achieved among different models.}
\label{tab:discuss_ml}
\small
\resizebox{\textwidth}{!}{
\begin{tabular}{clcrrrrr}
\toprule
\multicolumn{1}{l}{} &
  &
  \multicolumn{1}{l}{} &
  \multicolumn{1}{l}{\textit{\textbf{Gerrit}}} &
  \multicolumn{1}{l}{\textit{\textbf{GitHub}}} &
  \multicolumn{1}{l}{\textit{\textbf{Google Play}}} &
  \multicolumn{1}{l}{\textit{\textbf{Jira}}} &
  \multicolumn{1}{l}{\textit{\textbf{StackOverflow}}} \\
  \hline
  &  & \textbf{ALBERT} & 0.53 & 0.64 & 0.42 & 0.84 & 0.43 \\
  &  & \textbf{BERT} & 0.52 & 0.65 & \cellcolor[HTML]{F4CCCC}0.58 & 0.88 & 0.53 \\
  &  & \textbf{DistilBERT} & 0.51 & 0.64 & 0.56 & 0.81 & 0.44 \\
  &  & \textbf{RoBERTa} & 0.46 & 0.6 & 0.29 & 0.62 & 0.41 \\
  & \multirow{-5}{*}{\bf Naive Bayes} & \textbf{XLNet} & 0.52 & 0.41 & 0.46 & 0.74 & 0.29 \\
  \cline{2-8}
  &  & \textbf{ALBERT} & 0.63 & 0.56 & 0.42 & 0.89 & 0.48 \\
  &  & \textbf{BERT} & 0.5 & 0.61 & 0.54 & 0.82 & 0.37 \\
  &  & \textbf{DistilBERT} & 0.52 & 0.56 & 0.48 & 0.88 & 0.33 \\
  &  & \textbf{RoBERTa} & 0.57 & 0.55 & 0.48 & 0.75 & 0.31 \\
  & \multirow{-5}{*}{\bf Decision Tree} & \textbf{XLNet} & 0.58 & 0.44 & 0.25 & 0.68 & 0.4 \\
  \cline{2-8}
  &  & \textbf{ALBERT} & 0.49 & 0.73 & 0.37 & 0.92 & 0.3 \\
  &  & \textbf{BERT} & 0.46 & 0.73 & 0.52 & \cellcolor[HTML]{F4CCCC}0.97 & 0.3 \\
  &  & \textbf{DistilBERT} & 0.53 & 0.76 & 0.51 & 0.92 & 0.39 \\
  &  & \textbf{RoBERTa} & 0.43 & 0.74 & 0.48 & 0.94 & 0.3 \\
  & \multirow{-5}{*}{\bf Random Forest} & \textbf{XLNet} & 0.46 & 0.54 & 0.3 & 0.78 & 0.3 \\
  \cline{2-8}
 \multirow{-16}{*}{\bf Macro-F1} & {\bf Fine-tuning} & \textbf{Best} & \cellcolor[HTML]{F4CCCC}0.81 & \cellcolor[HTML]{F4CCCC}0.94 & 0.57 & \cellcolor[HTML]{F4CCCC}0.97 & \cellcolor[HTML]{F4CCCC}0.68 \\
 \hline
  &  & \textbf{ALBERT} & 0.55 & 0.64 & 0.51 & 0.86 & 0.57 \\
  &  & \textbf{BERT} & 0.56 & 0.66 & 0.69 & 0.89 & 0.65 \\
  &  & \textbf{DistilBERT} & 0.54 & 0.65 & 0.66 & 0.83 & 0.54 \\
  &  & \textbf{RoBERTa} & 0.47 & 0.63 & 0.43 & 0.64 & 0.49 \\
  & \multirow{-5}{*}{\bf Naive Bayes} & \textbf{XLNet} & 0.54 & 0.48 & 0.51 & 0.78 & 0.49 \\
  \cline{2-8}
  &  & \textbf{ALBERT} & 0.72 & 0.57 & 0.57 & 0.91 & 0.71 \\
  &  & \textbf{BERT} & 0.65 & 0.62 & 0.6 & 0.84 & 0.68 \\
  &  & \textbf{DistilBERT} & 0.62 & 0.56 & 0.69 & 0.89 & 0.66 \\
  &  & \textbf{RoBERTa} & 0.68 & 0.56 & 0.66 & 0.79 & 0.64 \\
  & \multirow{-5}{*}{\bf Decision Tree} & \textbf{XLNet} & 0.68 & 0.46 & 0.34 & 0.71 & 0.7 \\
  \cline{2-8}
  &  & \textbf{ALBERT} & 0.75 & 0.74 & 0.57 & 0.93 & 0.8 \\
  &  & \textbf{BERT} & 0.75 & 0.74 & 0.74 & \cellcolor[HTML]{F4CCCC}0.97 & 0.8 \\
  &  & \textbf{DistilBERT} & 0.77 & 0.77 & 0.74 & 0.93 & 0.82 \\
  &  & \textbf{RoBERTa} & 0.75 & 0.76 & 0.71 & 0.95 & 0.8 \\
  & \multirow{-5}{*}{\bf Random Forest} & \textbf{XLNet} & 0.75 & 0.59 & 0.49 & 0.84 & 0.8 \\
  \cline{2-8}
 \multirow{-16}{*}{\bf Micro-F1} & {\bf Fine-tuning} & \textbf{Best} & \cellcolor[HTML]{F4CCCC}0.86 & \cellcolor[HTML]{F4CCCC}0.94 & \cellcolor[HTML]{F4CCCC}0.83 & \cellcolor[HTML]{F4CCCC}0.97 & \cellcolor[HTML]{F4CCCC}0.89 \\
\hline
  &  & \textbf{ALBERT} & 0.62 & 0.79 & 0.69 & 0.91 & 0.64 \\
  &  & \textbf{BERT} & 0.57 & 0.83 & 0.73 & 0.93 & 0.81 \\
  &  & \textbf{DistilBERT} & 0.57 & 0.79 & 0.76 & 0.91 & 0.74 \\
  &  & \textbf{RoBERTa} & 0.59 & 0.79 & 0.6 & 0.75 & 0.69 \\
  & \multirow{-5}{*}{\bf Naive Bayes} & \textbf{XLNet} & 0.62 & 0.65 & 0.68 & 0.8 & 0.56 \\
  \cline{2-8}
  &  & \textbf{ALBERT} & 0.63 & 0.67 & 0.59 & 0.89 & 0.63 \\
  &  & \textbf{BERT} & 0.5 & 0.71 & 0.64 & 0.84 & 0.53 \\
  &  & \textbf{DistilBERT} & 0.52 & 0.68 & 0.64 & 0.88 & 0.5 \\
  &  & \textbf{RoBERTa} & 0.57 & 0.67 & 0.65 & 0.75 & 0.48 \\
  & \multirow{-5}{*}{\bf Decision Tree} & \textbf{XLNet} & 0.58 & 0.58 & 0.43 & 0.69 & 0.56 \\
  \cline{2-8}
  &  & \textbf{ALBERT} & 0.73 & 0.89 & 0.76 & 0.99 & 0.72 \\
  &  & \textbf{BERT} & 0.68 & 0.91 & 0.77 & 0.996 & 0.69 \\
  &  & \textbf{DistilBERT} & 0.65 & 0.9 & 0.84 & 0.99 & 0.68 \\
  &  & \textbf{RoBERTa} & 0.64 & 0.9 & 0.79 & 0.98 & 0.72 \\
  & \multirow{-5}{*}{\bf Random Forest} & \textbf{XLNet} & 0.67 & 0.76 & 0.67 & 0.85 & 0.63 \\
  \cline{2-8}
\multirow{-16}{*}{\bf AUC} & {\bf Fine-tuning} & \textbf{Best} & \cellcolor[HTML]{F4CCCC}0.89 & \cellcolor[HTML]{F4CCCC}0.99 & \cellcolor[HTML]{F4CCCC}0.89 & \cellcolor[HTML]{F4CCCC}0.998 & \cellcolor[HTML]{F4CCCC}0.87 \\
\bottomrule
\end{tabular}
}
\end{table}

\textcolor{black}{
\subsection{Additional Experiments}
In addition to our primary experiments, we carried out one supplementary set of experiments to further enrich and substantiate our study.
}

\textcolor{black}{\textbf{An Alternative Way to Leverage sLLMs.}}
\textcolor{black}{
Building on existing research~\cite{zhang2020sentiment}, we adapt sLLMs for sentiment label prediction. An alternative approach involves leveraging sLLMs' embeddings in conjunction with various machine learning (ML) classifiers. Our study investigates the efficacy of this methodology. Specifically, we evaluate five sLLMs, as discussed in Section~\ref{sec:evaluated_lms}, in conjunction with three distinct ML classifiers: Naïve Bayes~\cite{rish2001empirical}, Decision Tree~\cite{priyam2013comparative}, and Random Forest~\cite{breiman2001random}. These ML models are implemented using the scikit-learn library~\cite{scikit-learn}, utilizing their default settings.
We take the first token ([CLS]) embedding for each document in our dataset as document representation.
}

\textcolor{black}{
The results presented in Table~\ref{tab:discuss_ml} offer a comprehensive comparison between the efficacy of fine-tuning sLLMs and employing ML models with static sLLM embeddings. It is observed that, in most cases, fine-tuning sLLMs leads to superior macro-F1 scores compared to traditional ML classifiers with fixed embeddings. A notable exception to this trend is observed with the Naïve Bayes classifier on the \gp dataset.
In our evaluation of ML classifiers, Random Forest stands out, achieving the highest average macro-F1 score of 0.57, micro-F1 score of 0.77, and AUC value of 0.79 across various datasets and embedding models. In comparison, the approach of fine-tuning sLLMs results in an average macro-F1 score of 0.75, which is a significant 32.3\% improvement over the Random Forest model. Additionally, fine-tuning sLLMs achieves an average micro-F1 score of 0.85, surpassing the Random Forest model by 10.5\%. Fine-tuning sLLMs can achieve an average AUC value of 0.89, which is 12.4\% higher than the Random Forest model. This demonstrates the considerable advantages of fine-tuning sLLMs in terms of all the macro-F1, micro-F1, and AUC scores.
}

\subsection{Implications for Future Research}
Based on the experimental results in our study, we derive the empirical guidelines for future research on SA4SE and SE in general as follows.

\vspace{8px}
\noindent{\textbf{Select your approach: considering the size and class distribution of the dataset.}}
When determining the most suitable strategy for a specific task, it is important to consider the size and class distribution of the dataset. Based on our empirical results, it is crucial to recognize that, in cases with ample and balanced training data, fine-tuning sLLMs remains the preferred choice. This guideline is applicable to numerous SE tasks.
If there are already manually curated datasets available or acquiring labeled data is not a significant challenge, fine-tuning sLLMs represents a straightforward and effective option. 
However, in scenarios where labeled datasets are scarce, bLLMs emerge as a potential solution.

\vspace{8px}
\noindent{\textbf{Effective prompt engineering unlocks the full potential of bLLMs.}}
Our experiments reveal a crucial insight: while prompt templates may appear similar at first glance, determining which one will yield the highest accuracy requires actual execution and template refinement. 
\textcolor{black}{Carefully crafting prompt templates proves advantageous, particularly in zero-shot scenarios, as it allows us to better leverage the capabilities of bLLMs. For future research, we recommend testing multiple prompt templates on a subset of samples and selecting the best-performing ones for deployment. This meticulous process of prompt engineering and evaluation is essential for optimizing the performance of bLLMs. Furthermore, our findings suggest that in few-shot learning scenarios, adding more examples (shots) is not always beneficial. Depending on the task characteristics, a longer context can potentially confuse bLLMs and lead to suboptimal results. Therefore, researchers should carefully evaluate the impact of increasing the number of shots, as it may not consistently improve performance.
}

\vspace{8px}
\noindent{
    \textbf{Comparison of prompt engineering for LLMs in the SA4SE task and other SE tasks.} The SA4SE task is relatively straightforward, with the primary goal of determining the sentiment of a given text from SE artifacts. This simplicity allows for the creation of concise and effective prompt templates. Thus, this study focuses on investigating whether minor prompt changes would lead to significant performance differences in the models.
    However, some other SE tasks, such as code summarization~\cite{ahmed2024automatic}, bug replay~\cite{feng2024prompting}, and vulnerability detection~\cite{zhou2024large}, often contain additional contextual information that can be included in the prompt. For instance, the prompt used for code summarization~\cite{ahmed2024automatic} includes repository information, variable names, and scopes. Such information is unavailable in the SA4SE task, where the prompt is limited to the text to be analyzed. These task-specific details can guide the model in generating the desired output. In these SE tasks, prompt engineering may focus on which pieces of information to include in the prompt to guide the model's output generation.
    In contrast, for SA4SE, prompt engineering may focus more on word choice and whether to include additional examples in the prompt. Given the diversity of SE tasks, future research should explore developing task-specific prompt templates to maximize the performance of LLMs. This tailored approach can enhance the adaptability of LLMs to various SE tasks, ensuring optimal performance across different domains.
}

\vspace{8px}
\noindent{\textbf{Additional guidelines for rule setting in human-labeled SA datasets or prompt design.}
As elaborated in Section~\ref{sec:error}, the inconsistency in labeling practices across various datasets poses a significant challenge for bLLMs to predict labels accurately. To enhance performance, we propose two potential approaches:
1. Encouraging human annotators to adhere to general labeling rules when annotating data.
2. Empowering bLLMs to incorporate dataset-specific labeling rules.
Recall that despite restricting the label options to ``positive'' and ``negative'', bLLMs still occasionally return ``neutral'' labels when assessing the \jr dataset. This observation underscores the importance of further exploring and refining prompts to match the dataset characteristics.
}

\subsection{Threats to Validity}

\noindent{\textbf{Threats to Internal Validity.}} One potential source of bias in our empirical study may arise from the choice of prompt templates. To address this concern, we conducted experiments in the zero-shot setting using three different prompt templates. \textcolor{black}{It is important to note that we based our prompt templates on prior work~\cite{zhang2023sentiment}. Additionally, we examined the influence of the number of shots in the few-shot setting.} 
Furthermore, there is a potential concern regarding the quality of the labeled dataset.
We did not generate new datasets but rather relied on pre-existing ones from other sources.
Consequently, we inherit this quality concern from the original works.
However, as described in Section~\ref{sec:dataset}, in the original labeling process, each dataset was labeled by two or more labelers labeled individually and resolved the conflict by involving another labeler.
Thus, we consider the threat to be minimal.

\vspace{8px}

\noindent{\textbf{Threats to External Validity.}}
Our findings may not necessarily generalize to data from other platforms. Nevertheless, we have taken steps to mitigate this threat by considering data from five distinct platforms. It is important to recognize that our results are specific to the dataset and experimental setup we employed. In the few-shot learning setting, our results are contingent on randomly sampled examples. Nevertheless, our experiments and results still offer valuable insights, demonstrating that bLLMs can be a promising approach when dealing with a scarcity of annotated data.
In the future, we plan to expand our analysis by incorporating additional datasets from various platforms and exploring more diverse prompt templates to enhance our understanding of leveraging bLLMs for SE in SE further.

\textcolor{black}{
Another potential threat is our reliance on a single held-out test set for evaluation. While this approach is common in the SA4SE literature~\cite{novielli2018benchmark,calefato2018sentiment,novielli2020can,zhang2020sentiment}, it may not fully capture the diversity of scenarios and data distributions that the models are intended to handle. Many other SE tasks utilizing bLLMs also adopt the held-out test set approach, such as code comment generation~\cite{geng2024large}, log parsing~\cite{ma2024llmparser}, and automatic logging~\cite{xu2024unilog}.
Beyond the SE research areas, other domains, including natural language processing and information retrieval, have also employed the held-out setting for evaluation, e.g., stream recommendation~\cite{zhang2024gpt4rec}, document-level sentiment analysis~\cite{song2023sequence}, generative outfit recommendation~\cite{xu2024diffusion}.
Although cross-validation may provide a more comprehensive assessment of performance across different situations, we believe the held-out test set approach is sufficient for our study, given its wide adoption in the literature. In the future, we plan to incorporate cross-validation into our evaluation methodology to enhance the robustness and reliability of our findings.
}

\textcolor{black}{The additional potential threat to external validity is the chosen bLLMs can be surpassed by newer models. Despite this, our research is still relevant as it offers crucial insights into the current capabilities and constraints of the accessible open-source bLLMs. It establishes a benchmark for evaluating the progression of bLLMs over time. By documenting the specific versions of the models utilized, we guarantee the reproducibility of our findings. This commitment ensures that our work serves as a foundational reference, facilitating subsequent research and exploration in the evolving domain of bLLMs on the SA4SE task.}

\vspace{8px}
\noindent{\textcolor{black}{\textbf{Threats to Construct Validity.} The first consideration for the construct validity of our study is the potential for data leakage, particularly given our use of sLLMs and bLLMs. Data leakage could occur if the evaluation dataset was already present in the pre-training dataset of the models, potentially inflating model performance. However, we assess this risk to be low in our study. The dataset we employed is not readily accessible through standard web browsing and requires downloading from a specific URL. This aspect reduces the likelihood that it was included in the LLM's pre-training data, mitigating the risk of data leakage affecting our results.
}}

Another potential threat is the unreliable labeled data. As discussed in Section~\ref{sec:error}, a significant challenge bLLMs encountered was the inconsistency in labeling rules and guidelines within the ground truth dataset. This inconsistency poses a threat to the construct validity of our study, as it can lead to confusion in model training, potentially resulting in suboptimal performance. To address this issue and enhance the reliability of our findings, future research should focus on establishing clear, consistent labeling rules prior to the creation of new datasets. Additionally, a thorough manual review of existing labeled datasets is recommended to ensure uniformity and accuracy, thereby mitigating the risks associated with inconsistent labeling practices.

\textcolor{black}{In addition, our choice to label ``neutral'' as the opposite of the ground truth in the \jr dataset introduces an additional risk. This approach was adopted to establish a lower-bound for performance, which might, however, compromise the effectiveness of bLLMs. Similarly, we also map non-explicit responses from bLLMs as ``neutral'' poses a potential threat. This decision, aimed at ensuring consistency in SA, is based on the rationale that ``mixed'' sentiments typically indicate a balanced or uncertain position, closely aligning with a \textit{neutral} stance.}

\textcolor{black}{Lastly, to obtain prediction probability scores from bLLMs, we directly prompt the models for their probability scores for each label. In cases where the total sum of these scores does not equal 1, we normalize them by dividing each score by the total sum. However, this approach does not extract the probability scores directly from the model's internal computations, which may not be the most optimal method for obtaining accurate probability scores. This could potentially introduce noise into our results, as the probability scores may not align with the actual probabilities.
For instance, in the \gp dataset, the \llama model made perfect predictions on the test set with 5-shot learning, but the AUC value was only 0.76. In one case, although the correct prediction label was negative, the predicted probability for the negative label was only 0.2, while the probability for the positive label was 0.8. While such cases are rare, we believe this threat is minimal. Moreover, our main evaluation metric is macro-F1, which is calculated based on the precision and recall of each class and does not consider the probability scores of the models. Thus, our study is not highly affected by potential inaccuracies in the AUC metric.
To mitigate this threat, we plan to explore alternative methods for extracting probability scores from bLLMs in future research.}
\section{Conclusion and Future Work}
\label{sec:conclusion}
In conclusion, our study marks the initial step towards comprehending the potential of utilizing prompting bLLMs in discerning sentiment within SE domain documents. Our experiments reveal that in cases with limited annotated data, bLLMs outperform sLLMs, and zero-shot learning is a viable approach. However, when substantial and well-balanced training data is available, fine-tuning sLLMs is the preferable strategy over prompting bLLMs. 

Looking ahead, our future work will explore the versatile applications of bLLMs to enhance their efficacy in the SA4SE task. 
Meanwhile, we plan to apply the latest advancements in transformer accelartion~\cite{duglide,du2021order} to speed up the training and inference of bLLMs.
We also plan to leverage the results of SA4SE for other downstream tasks, such as library or API recommendations~\cite{zhang2022benchmarking,irsan2023multi}.

\vspace{8px}
\noindent{\textbf{Replication Package.}} We release the data, code and results on: \url{https://github.com/soarsmu/LLM4SA4SE}.

\section*{Acknowledgments}
This research / project is supported by the National Research Foundation, under its Investigatorship Grant (NRF-NRFI08-2022-0002). Any opinions, findings and conclusions or recommendations expressed in this material are those of the author(s) and do not reflect the views of National Research Foundation, Singapore.

\bibliographystyle{ACM-Reference-Format}
\bibliography{main}

\end{document}